\documentclass[aps,onecolumn,pra,12pt]{revtex4-2}
\usepackage{braket,graphicx}
\usepackage{enumerate}
\usepackage{xcolor}
\usepackage{amsmath,amsfonts,amssymb,amsthm}
\usepackage{nicefrac,bm,mathrsfs,color,dsfont}
\newcommand{\tcr}{\textcolor{red}}
\begin{document}
\title{Quantum Computation}
\author{Barry C. Sanders\href{https://orcid.org/0000-0002-8326-8912}{\includegraphics[scale=0.05]{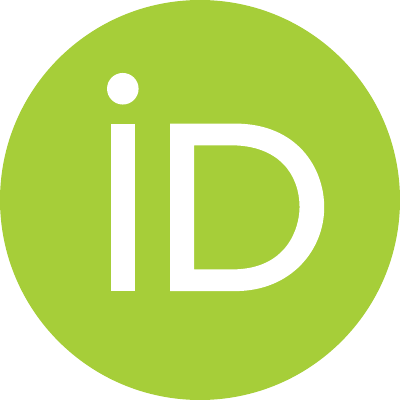}}}
\email{sandersb@ucalgary.ca; +1-403-210-8462}
\affiliation{Institute for Quantum Science and Technology, University of Calgary,\\
2500 University Drive NW, Calgary, Alberta, Canada T2N 1N4}

\date{\today}
\begin{abstract}
This chapter summarizes quantum computation, including the motivation for introducing quantum resources into computation and how quantum computation is done.
Finally, this chapter articulates advantages and limitations of quantum computation, both fundamental and practical.
\end{abstract}

\maketitle

\section{Introduction}
Computation is about transforming input information to output information~\cite{Sip12}.
Information comprises symbols as an alphabet,
usually represented by finite strings of binary digits (bits)
$\bm{b}=(b_k)_{k=0}^{n-1}\in\{0,1\}^n$,
but could be distributions thereof~\cite{ZSS21},
although radically different informational foundations such as real numbers $\mathbb R$~\cite{BSS89} are possible.
Informational transformations are governed by a well defined logical process,
founded on the basic logical steps inherent in the physical realisation of the information-processing machine
such as a Turing machine,
which serves as a model for deterministic computation based on recursive functions~\cite{Tur36}.
These logical processes could be about solving equations~\cite{Tho76},
or executing an algorithm according to basic arithmetic or logical operations on a computer, which is a programmable machine
capable of executing sequences of such operations,
or instructions.
The framework of computer science separates computational problems into types
that include decision, function, search, optimisation, and feasibility problems.
Problems can be computable or not~\cite{R62} and have different complexity~\cite{Gol08}, which are studied in theoretical computer science.

As information is physical~\cite{Lan61,Ben73},
treating the Turing machine as a physical device is consequently convenient,
and then quantum computation enters by quantizing the classical physics describing that machine~\cite{Deu85}.
Superpositions of information states are thus possible,
and ideal processing of such quantum information states can be described by unitary transformations,
followed by quantum measurements with their probabilistic nature.
The logical foundation differs for quantum computation compared to classical (i.e., using non-quantum physics) computation---Boolean logic is replaced by quantum logic.

This chapter focuses on quantum computation,
with computational problems,
and therefore input and output, being classical, but the computational processing of this information being in the quantum domain, hence involving quantum logic operations.

\section{Boolean and reversible computation}
\label{sec:Booleanreversible}

Computation is conveniently understood in terms of Boolean algebra,
which formalises processing logical propositions mathematically.
A $k$-ary Boolean function
$f:\{0,1\}^k\to\{0,1\}:
    \bm{b}\mapsto f(\bm{b})$
represents a decision problem,
also known as a (formal) language,
with $k=1$ for unary and $k=2$ for binary functions.
The unary functions are identity~$\mathds1$,
which leaves the bit unchanged,
and NOT, or `negation', which flips:
$\neg b\gets b\oplus1$
for~$\oplus$ summation modulo~2,
$\lnot$ for negation
and~$\gets$ for `gets'.

In Boolean logic, a literal comprises both a logical variable $b\in\{0,1\}$ and its negation~$\neg b$.
Binary functions include~$\oplus$ for `exclusive or'
and~$\land$ (conjunction) and~$\lor$ (disjunction)
for logical `and' and `or' operations,
respectively. A 
Boolean function~$f$ is decidable if the output is true~($0$) conditioned on the proposition itself is true,
and false~($1$)
conditioned on the proposition itself is false.
Ideally,
a program accepts input and delivers an output representing the solution to the problem in a finite number of steps and then halts.
However, some problems, such as the halting problem~\cite{Kle52} are undecidable
(not every problem corresponds to a program that halts after a finite number of steps),
hence uncomputable, because either of these conditions fails.

Consider any $n$-bit Boolean function~$f$.
Remarkably,
a $k$-ary value $K\geq2$ can be fixed such that~$f$
can be solved via a composition of $k$-ary
Boolean functions with $k\leq K$
functions independent of the choice of~$n$.
This universal set of $k\leq K$ generating operations is the `instruction set',
and the composition of these $k$-ary functions is the `circuit' to solve~$f$.
Moreover,
certain $k$-ary functions,
such as NAND ($\neg\circ\land$ with~$\circ$ signifying composition of functions),
are universal in the sense that compositions of this binary function alone generates any circuit.

From Boolean functions,
we can build the notion of Boolean circuits.
A Boolean circuit is a finite directed acyclic graph~\cite{BG09},
whose vertices comprise~$n$ input bits and a universal set of Boolean gates
(with the number of such gates being the circuit size)
and has output bits.
The Boolean circuit can be designed to compute Boolean function~$f$,
which implies just one output bit,
and the length of the circuit is the maximum number of edges to go from an input to an output gate.
Circuit-depth complexity for~$f$ is the minimum of these maximum lengths over all Boolean circuits designed to compute~$f$.

Time in computation is related to circuit depth.
Measuring time in this way is valuable for studying computational complexity as this time depends on the logical operations and not on how fast the physical hardware, such as the speed of a particular implementation of NAND, is.
If a Turing machine computes~$f$
in time~$T(n)$,
then a Boolean circuit can solve~$f$
in depth~$\tilde{O}(T(n))$~\cite{PF79},
meaning the same time scaling up to logarithmic factors with~$\tilde{O}$
as explained in Eq.~(\ref{eq:softO}).

The space resource of a circuit refers to the number of bits required for the logical circuit.
Time-space trade-offs exist such that more or less space could be used to decrease or increase, respectively, the computational time as circuit depth.
Resource consumption is quantified by the computational \emph{cost},
which need not be unique.

A computational problem is defined over instances;
for example the problem of factoring semiprime numbers (numbers expressible as $pq$
with $p,q\in\mathbb{P}$ for~$\mathbb{P}$ the set of prime numbers) accepts different semiprime instances such as~$3763=53\times71$ and~$5141=53\times97$.
Computational complexity studies
how the (usually worst-case) cost of solving the problem increases as the size of the instance,
measured as the minimum number~$n$ of bits to specify the instance,
grows.

For studies of asymptototic computational complexity,
resource scaling in the large-instance limit is important.
Bachmann-Landau notation~\cite{CLRS09} is used, 
with~$O(g(x))$ defined by
\begin{equation}
\label{eq:bigO}
\left\{f(x)|
\exists M>0,x_0\in\mathbb{R}:
|f(x)|\leq Mg(x)\forall x\geq x_0\right\}.
\end{equation}
We define
\begin{equation}
\label{eq:softO}
\text{poly}x:=\bigcup_{c\in\mathbb{R}^+} O(x^c),\,
\tilde{O}(f(x))
    :=O\left(f(x)\text{polylog}x\right)
\end{equation}
with the latter known as soft-O notation that
incorporates polynomial-of-logarithm factors
with polylog being `poly' over `log'.

Complementary to big-O~(\ref{eq:bigO}),
$o(g(x))$ is defined by
\begin{equation}
\left\{f(x)|\forall M > 0 \exists a\in\mathbb{R}: 0
\leq f(x)<Mg(x)\forall x>a\right\}
\end{equation}
meaning that~$f(x)$ is smaller than~$g(x)$ for large~$x$.
For lower bounds~\cite{Knu76},
the relation
\begin{equation}
g(x)\in\Omega(f(x))\Leftrightarrow f(x)\in O(g(x)),
\end{equation}
which defines~$\Omega$,
is useful, and
\begin{equation}
\label{eq:tightbound}
\Theta(f(x)):=O(f(x))\cap\Omega(f(x))
\end{equation}
expresses tight bounds.
With respect to computational complexity,
the number of bits~$n$ replaces~$x\in\mathbb{R}$ in these expressions,
and we use Bachmann-Landau notation~\cite{CLRS09} in discussing complexity of classical and quantum algorithms and classical here.

A decision problem, or language~$L$,
is deemed to be tractable if both space and time costs scale as poly$n$
by Cobham's Thesis~\cite{Cob65};
i.e., $L$ is in complexity class~\textsf{P}
meaning polynomial (henceforth `poly' for short) time on a Turning machine;
\textsf{P}$\subseteq$\textsf{PSPACE},
which means poly space (memory),
and \textsf{PSPACE} contains all complexity classes within what is known as the polynomial-time hierarchy~\cite{Sto76}.
Conversely, $L$ is `hard'
if all known algorithms are superpoly$n$,
i.e., cannot be bounded in poly time so `superpolynomial' in the worst-case scenario.

As a special case of hard problems,
the complexity class~\textsf{NP}
refers to nondeterministic poly-time,
meaning~$L$ is solved in poly time on a \emph{nondeterministic} Turing machine.
A nondeterministic Turing machine conducts a brute-force search
by allowing a multitude of actions in parallel
in contrast to the deterministic case with fixed action,
and accepting the input if \emph{any} branch of actions halts with an accept condition.
The probabilistic Turing machine is a special case of a non-deterministic Turing machine wherein a probability distribution is provided to make weighted stochastic choices of computational transitions at each state.

Tractable computation on a nondeterministic Turing machine is equivalent to being computationally hard to solve but easy to verify on a (deterministic) Turing machine.
Obviously, \textsf{P}$\subseteq$\textsf{NP},
but whether \textsf{P}$=$\textsf{NP} is a famous open problem~\cite{For09}.
With respect to any complexity class~\textsf{C},
$L_0$ is \textsf{C-Hard} if every $L\in$\textsf{C} can be reduced in poly time to~$L_0$,
and~$L$ is \textsf{C-complete}
if $L_0$ is both \textsf{C-hard} and a member of~\textsf{C}. 
 \textsf{NP-complete} is an important example
relevant to combinatorial optimisation~\cite{LLRS91}
and to approximate optimisation~\cite{SG76}.

The probabilistic Turing machine model deliberately leads to erroneous results:
sometimes accepting strings that should be rejected and vice versa.
Formally,
$L$ is \emph{recognised} by the probabilistic Turing machine if,
given probability~$\varepsilon$,
a string in~$L$ and a string not in~$L$
are accepted or rejected with probability $1-\varepsilon$,
respectively.
The complexity class \textsf{BPP}
corresponds to languages for $\varepsilon=\nicefrac13$
(and \textsf{PP},
or probabilistic polynomial-time,
for any $\varepsilon<\nicefrac12$~\cite{Gil77}, which contains~\textsf{NP}).
Obviously,
\textsf{P}$\subseteq$\textsf{BPP},
but whether the widely believed equality \textsf{P}$=$\textsf{BPP} is true is an open problem.

The Boolean satisfiability problem (SAT)---deciding the 
existence of an assignment of values to a Boolean expression of literals such that the output is true---is an important type of computation~\cite{Vaz01} with special significance for quantum computation~\cite{Bra11}.
Expressed in conjunctive normal form,
$k$-SAT is expressed as a 
conjunction ($\land$) of fixed-length clauses
with each clause
a disjunction ($\lor$) of~$k$ literals.
Not only is 2-SAT$\in$\textsf{P}~\cite{Kro67},
but 2-SAT$\in$\textsf{NL-complete}~\cite{GJ79} as well,
with~\textsc{NL} meaning use of a logarithmic amount of space
on a nondeterministic Turing Machine.
In contrast,
3-SAT$\in$\textsf{NP-complete}
and is pertinent for combinatorial optimization
as finding the exact optimal solution is typically \textsf{NP-complete}.

Reversible computation~\cite{Ben00} provides a valuable bridge from concepts of computation based on Boolean logic to concepts of quantum computation~\cite{Tof80,Per13}.
One impetus for reversible computation arises from Maxwell's demon,
which enables studying entropy and the second law of thermodynamics~\cite{MNV09}.
From Szilard's information engine incorporating Maxwell's demon~\cite{RC20},
Landauer connected information to thermodynamics by showing
that erasure is physical~\cite{Lan61}.

Reversible computation shows
that bit erasure can be postponed so the full computation can be achieved in a logically reversible way
with the solution plus the `garbage' bits at the output potentially serving as the input with the reversed version of the computation mapping this solution to the original input.
For a finite machine undergoing a large number of computations,
erasures are eventually needed to recoup space from garbage for useful information.
However, energy consumption requirements per computational step can approach zero in a reversible computer.

Reversible computation is achieved by embedding
the Boolean function~$f$
into a reversible function as
\begin{align}
\label{eq:reversiblefunction}
F:\{0,1\}^{n+1}\to\{0,1\}^{n+1}
:(\bm{b},b')\mapsto(\bm{b},b'\oplus f(\bm{b}))
\end{align}
for $\{0,1\}^{n+1}:=\{0,1\}^n\times \{0,1\}$,
making reversible computation equivalent to permutation~\cite{CJS22}.
Analogous to Boolean logic,
$F$ can be decomposed into a sequence of reversible $k$-ary logic gates,
with~$k$ independent of the choice \tcr{of~$n$.}
Whereas the binary NAND gate suffices as a universal gate for Boolean logic,
a ternary gate, mapping three bits to three bits,
is the smallest value for~$k$ in reversible logic.

The Toffoli, or controlled-controlled-not gate (CCNOT) is one such universal gate~\cite{Tof80}
with C$^{\iota}$NOT
mapping~$\iota$ control bits and one target bit to the same state unless all control bits are~$1$ in which case the target bit flips.
Another universal gate is the Fredkin~\cite{FT82}, or controlled-SWAP, gate~\cite{SK79}.
Thus,
following the language established for Boolean logic,
each of CCNOT and CSWAP is a cardinality-one universal instruction set
for reversible computation.
As for Boolean logic,
decomposing logical circuits into the same kind of circuit allows careful comparison on relative complexity between different reversible functions~$F$.
\section{Quantum fundamentals}
\label{sec:foundationsqcomputation}
We begin with a brief mathematical description of nonrelativistic quantum mechanics~\cite{Str08}.
A state is a normalised positive linear functional on a unital C$^*$ algebra
(comprising linear operators on a Hilbert space
with the~$^*$ operation representing adjoint, typically~$^\dagger$ in physics).
Representing linear operators on Hilbert space is achieved by the Gel’fand-Naimark-Segal construction~\cite{GN43,Seg47},
and each state can be uniquely identified with a positive semi-definite trace-class 
(`density') operator of unit trace by requiring equivalence between expectation values for representations of operators
and the functional acting on the operator.

In the case of quantum bits
(`qubits'~\cite{Sch95}),
states are density operators on a tensor product of two-dimensional Hilbert spaces:
$\mathscr{H}_2^n$.
Pure states are idempotent operators and thus can represented by vectors in~$\mathscr{H}_2^n$,
which is isomorphic to projective complex space of dimension $2n-1$,
i.e., $P_{2n-1}(\mathbb{C})$,
with~$2n$ the overall Hilbert-space dimension and subtraction of one for the constraint on norm.
Each multiqubit state is thus representable as a complex unit-length vector of length~$2n$.

In Dirac notation,
a qubit string is
\begin{equation}
\label{eq:multiqubitstate}
\ket\psi^n=\sum_{\bm{b}\in\{0,1\}^n}\psi_{\bm{b}}\ket{\bm{b}},\,
\left(\psi_{\bm{b}}\right)\in P_{2*-1}(\mathbb{C}),
\end{equation}
with~$\{\ket{\bm{b}}\}$ the `computational basis states' of $\mathscr{H}_2^n$,
i.e., basis vectors labelled by bit strings.
Henceforth, we do not explicitly normalise the states if norm-$1$ is implied.
Two useful examples of non-product states are
`ebits'
(maximally entangled pair of qubits)
$\ket{0\;b_1}
    +(-1)^{b_0}\ket{1\;1-b_1}$.
The Hadamard gate maps
$H\ket{b}\mapsto\ket0+(-1)^b\ket1$,
and normalisation is implicit
in both cases.
Each multi-qubit gate is representable in~$\mathcal{M}_d(\mathbb{C})$,
meaning a $d\times d$ complex matrix,
where $d=2^n$ for~$n$ qubits.
The distance between two normalised multiqubit states~(\ref{eq:multiqubitstate})
is the `quantum angle'
$\arccos\left|\braket{\psi'|\psi}\right|$
corresponding to the Fubini-Study metric for~$\mathscr{H}$~\cite{BH01}.

On an ideal quantum computer,
$n$ qubits are transformed by unitary maps,
i.e., isometries on Hilbert space: $U\ket\psi^n$
with~$U\in\mathcal{M}_{2^n}(\mathbb{C})$ a representation of a group element in~SU$(2^n)$. 
The unitary map~$U$ is generated by a self-adjoint Hamiltonian operator~$H$ that describes the dynamical physical system yielding this gate,
and~$U$ forms a strongly continuous one-parameter unitary group over time~$t$.
The distance between two operators, say~$A$ and~$A'$ (need not be unitary),
is
$\text{dis}(A,A'):=\|A-A'\|$
for
$\|A\|:=\sup_{\left\|\ket\psi\right\|=1}
\left\|A\ket\psi\right\|$
the spectral norm~\cite{Wat18}.

Whereas Boolean and reversible-logic functions are exactly decomposable to a sequence of instruction gates,
with the universal instruction set having just one universal gate,
a related decomposition theorem for quantum computation~\cite{DN06}
yields a sequence that only approximately realizes~$U$
and requires at least two instruction gates.
The Solovay-Kitaev theorem,
as formulated by Dawson and Nielsen~\cite{DN06},
conveys that,
given a universal instruction set for SU$(d)$ and accuracy~$\epsilon>0$,
$\exists c>0$ such that,
$\forall U\in$SU$(d)$,
a sequence of instruction gates of depth~$O\left(\log^c\left(\nicefrac1\epsilon\right)\right)$
exists such that the unitary map so generated has a distance, based on operator norm,
not greater than~$\epsilon$.
We write $\log$ for logarithm in base~2 as the default.
Succinctly,
any universal instruction set generates a polylog$\left(\nicefrac1\epsilon\right)$-depth quantum circuit approximating~$U$,
and,
furthermore,
the classical algorithm to generate this sequence is also polylog$\left(\nicefrac1\epsilon\right)$,
with an even smaller constant~$c$
than for the circuit depth~\cite{DN06}.

A choice is made regarding which quantum instruction gates to include, 
with one common choice being~\cite{BBC+95}
\begin{equation}
\label{eq:HT}
\left\{\text{H}
=\begin{pmatrix}1&1\\1&-1\end{pmatrix}/\sqrt2,
\text{T}
:=\text{diag}(1,\omega_8),
\text{CNOT}\right\}
\end{equation}
for~$\omega_d$ the $d^\text{th}$ principal root of unity.
Alternatively,
combining~H with CCNOT 
whose representation is in~$\mathcal{M}_8\mathbb{C})$,
which is universal by itself for reversible computation,
provides a universal instruction set for quantum computation.
Physically,
the universal instruction set $\{\text{H},\text{T},\text{CNOT}\}$
is convenient as only single- and two-body interactions are involved,
whereas CCNOT involves more challenging three-body interactions.
\section{Quantum algorithms and heuristics}
\label{sec:quantumalgorithms}
In this section we consider quantum algorithms~\cite{Mos09}, quantum state generation and quantum heuristics,
which tend to be grouped together under the term `quantum algorithms'.
An algorithm is the means by which a computer solves a well posed computational problem and comprises
(a perhaps empty) input, an output and a procedure.
A procedure is expressed as a finite sequence of instructions based on a formal language and incurs finite cost.

On the other hand, heuristics are techniques for solving hard problems exactly or approximately,
such as for search or optimisation types of problems,
but are not guaranteed to succeed for finite cost.
Heuristics are valuable for solving \textsf{NP-hard} problems.
A metaheuristic is technique for devising or choosing or tuning a heuristic for attempting to solve optimisation or machine learning problems approximately~\cite{BDGG09},
with simulated annealing for approximate global optimisation one important example of a metaheuristic.

As discussed in \S\ref{sec:Booleanreversible},
reversible computation is permutation;
extending reversible computation to quantum computation leads to
permutations being replaced by unitary transformations.
The algorithmic procedure acting on a~$n$ qubits is then described by a unitary mapping~$U$ on~$\mathscr{H}_2^n$.
Approximate quantum computation is achieved by realising approximate~$U'$
such that dis$(U,U')<\epsilon$.
The approximation~$U'$
must be achievable as a quantum circuit 
with computational cost that grows as polylog$n$, i.e., efficient and known as `speedup'.
We  shall next discuss
important cases of quantum algorithms~\cite{Mon16,QAZ}.

Shor's quantum factoring algorithm, the most celebrated of all quantum algorithms,
was inspired by Simon's algorithm~\cite{Sim97}.
Simon's algorithm solves the following oracle problem,
with an oracle black box being a system
(e.g., a subroutine)
that accepts inputs and yields outputs with an unspecified procedure.
Cost is typically quantified by \emph{queries}
rather than in terms of time-space resources~\cite{Mos09}.
Given oracle $f:\{0,1\}^n\to\{0,1\}^n$,
subject to the promise that
$\exists s\in\{0,1\}^n$ such that
\begin{equation}
f(x)=f(y)\iff x\oplus y\in\left\{0^n,s\right\}
\forall x,y\in\{0,1\}^n,
\end{equation}
Simon's problem is to compute~$s$
with the fewest oracle queries possible
and can be converted to a decision problem by rewriting the task as distinguishing whether $s=0^n$ or not.
Importantly,
Simon's problem yields an exponential oracle separation between the two complexity classes,
namely \textsf{BPP}
and \textsf{BQP},
which is the quantum version of~\textsf{BPP}.

Simon's algorithm can be contrasted with the Bernstein-Vazirani algorithm~\cite{BV97},
which also proves a separation between \textsf{BPP}
and \textsf{BQP} but not exponentially.
The Bernstein-Vazirani algorithm 
accepts an oracle with input $x\in\{0,1\}^n$ and yields a single-bit output $x\odot h$
(with Hadamard multiplication notation,
which is element-wise vector or matrix multiplication)
with the task being to find the hidden string $h\in\{0,1\}^n$.
Quantumly,
a single query suffices~\cite{BV97};
classically, $O(n)$ queries to the oracle are required.

The Bernstein-Vazirani algorithm is itself
a specialisation of the Deutsch-Jozsa algorithm~\cite{DJ92}.
The Deutsch-Jozsa algorithm employs an oracle
that accepts an $n$-bit input and yields a single-bit output subject to the promise that all~$2^n$
possible outputs are either the same
%yield the same output 
(constant case) \emph{or}
half yield output~$1$ and the other half yield~$0$ (balanced case).
The task is to decide whether the oracle is constant or balanced,
which can be solved quantumly with just one query in contrast to two queries for the $n=1$ deterministic classical algorithm~\cite{Deu85}
and exp$n$ queries for general~$n$~\cite{DJ92};
However,
the task can be achieved with $O(1)$ queries with a probabilistic algorithm so the Deutsch-Jozsa algorithm yields an exponential speedup over \textsf{P}
but not over \textsf{BPP}.

Shor's algorithm delivers superpoly$n$ speedup for factoring any $n$-bit number~$N$
($n=\lceil\log N\rceil$)
compared to the best-known classical algorithms~\cite{Sho94,Sho97},
which shows that the computationally hard
problem of integer factoring
is tractable using quantum computation.
Specifically,
Shor's algorithm for factoring an $n$-bit integer~$N$ executes in $O(n^3)$ time~\cite{Sho94,Sho97},
subsequently reduced to $\tilde{O}(n^2)$~\cite{BCDP96},
which subexponentially outperforms the best known classical algorithm 
based on the general number sieve,
whose complexity is~\cite{Pom96}
\begin{equation}
\label{eq:gennumsieve}
O\left(\exp\left[c\left(\log n\right)^{\nicefrac13}\left(\log\log n\right)^{\nicefrac23}\right]\right)
\sim2^{\tilde{O}\left(n^{\nicefrac13}\right)}
\end{equation}
for~$c\in\mathbb{R}^+$,
and the soft-O
notation in the exponent on the right-hand side captures the log and log~log terms in the exponent on the left-hand side of Eq.~(\ref{eq:gennumsieve}).

Shor's algorithm comprises the following steps.
Given $N\in\mathbb{Z}^+$,
choose random $a\in[N]$%:=\{1,2,\dots,N\}$.
and compute $\gcd(a,N)$;
terminate if $\gcd(a,N)\neq1$
or else solve the order-finding problem,
namely
compute the smallest period~$r$
such that
\begin{equation}
\label{eq:periodfinding}
f_N(x+r)=f_N(x)
    \stackrel{\scriptscriptstyle N}=a^x.
\end{equation}
This order-finding problem is solved by Shor's quantum algorithm,
with $\stackrel{\scriptscriptstyle N}=$ 
meaning equal modulo~$N$.
Start over if the resultant minimum~$r$ is odd; else proceed.
If $a^{r/2}\stackrel{\scriptscriptstyle N}=-1$, start over;
else return $\gcd\left(a^{r/2}\pm1,N\right)$ and halt.
Shor's integer factoring algorithm dramatically affects cybersecurity,
particularly on breaking the Rivest-Shamir-Adleman (RSA) public-key encryption~\cite{RSA78}.
Minimising~$r$ is the classical computational bottleneck that leads to superpoly$n$ scaling for classical computation.

Shor's quantum period-finding algorithm works as follows.
Two input qubit strings with~$k$ source qubits and~$n$ target qubits
satisfy
\begin{equation}
N^2\leq2^k\leq2N^2,\,
n=\lceil N\rceil
\end{equation}
with $k$
chosen to ensure more than~$N^2$ terms in the sum.
Only~$r$ distinct values of~$a^q\bmod N$ are possible.

The first step is
\begin{equation}
H^{\otimes k}\ket0^{k+n-1}\ket1
    \mapsto
    \sum_{q=0}^{2^k-1}\ket{q}\ket0^{n-1}\ket1
\end{equation}
with the input state being all qubits in the~$\ket0$ except that the last target qubit is~$\ket1$.
An interesting fact is that the same transformation of $\ket0^k$
is achieved by the quantum Fourier transform (QFT)
\begin{equation}
\label{eq:qft}
\ket{q}\mapsto\sum_{q'=0}^{2^k-1}\omega_Q^{qq'}\ket{q'},
\end{equation}
following the convention of implicit state normalisation.
Note that the inverse QFT,
i.e., QFT$^{-1}$,
is obtained by replacing $\omega_Q^{qq'}$ with $\omega_Q^{-qq'}$.

The next step in the quantum circuit is to execute the quantum subcircuit for order-finding,
which can conveniently be viewed as an oracle problem.
The order-finding oracle yields the order~$r$ of a group 
in a full oracle-free circuit.
For some quantum algorithms discussed below,
the oracle approach and query cost is vital to showing quantum advantage,
but, for Shor's algorithm and its variants,
the oracle is replaced by an explicit quantum subcircuit and costs are expressed in space and time.

For quantum order-finding,
the oracle is replaced by a subcircuit that executes a quantum version of classical exponentiation by squaring~\cite{Gue12}.
This quantum subcircuit~\cite{MS12}
performs
$\ket{q,0}\mapsto\ket{q,a^q\bmod N}$
with~$\ket{q,0}$ having~$q$ as a label for the source qubit string and the second label~$0$ for the target qubit string.
Thus,
the output quantum state from the order-finding subcircuit is
\begin{equation}
\sum_{q=0}^{2^k-1}\ket{q,a^q\bmod N}
    =\sum_{p=0}^{r-1}\sum_{q=0}^{k-1}\ket{rq+p,a^p \bmod N},
\end{equation}
with this final form convenient to show how terms can be grouped together due to the modular exponent term in the target qubits.
Most of the quantum cost for Shor's algorithm is in executing this quantum subcircuit.

The next step of Shor's algorithm is to apply QFT$^{-1}$ to this output state,
which yields
\begin{equation}
\sum_{p=0}^{r-1} \sum_{q=0}^{r-1} \omega_r^{-p q}
\ket{k q,a^p \bmod N}.
\end{equation}
Measuring only the source qubit string~$s$ suffices to infer~$r$:
discarding the $s=0$ case,
$s$ is highly likely to be a multiple of~$k$
with coefficients from~$1$ to $r-1$.
An efficient number of repeats of this process suffices to estimate~$r$ with high confidence~\cite{San23}.

Quantum speedup for order finding also applies to other computational problems.
One case is the discrete logarithm problem,
posed as accepting $a,b,N\in\{0,1\}^n$ with the promise that
$\exists s:b\stackrel{\scriptscriptstyle N}=a^s$,
and the task is to find~$s$,
which can be achieved in poly$n$ time~\cite{Sho97}
compared to superpoly$n$ classically.
Quantum computers can solve the discrete logarithm problem on elliptic curves
using Shor's technique~\cite{Sho97},
thereby breaking elliptic curve cryptography~\cite{BL95,PZ03,MNV09}.

Another case concerns solving Pell's Equation
\begin{equation}
x^2-dy^2=1,\,
x,y,d\in\mathbb{Z},
\end{equation}
with~$d$ not a square.
The computational problem is 
\begin{equation}
(x_1,y_1)=\min \left(x+y\sqrt{d}:d\in\{0,1\}^n\right),
\end{equation}
which is uniquely identified by~$\lfloor R\rceil$
with $R:=\log(x_1+y_1\sqrt{d})$.
Whereas computing~$\lfloor R\rceil$ is superpoly$n$,
Hallgren's quantum algorithm finds~$\lfloor R\rceil$ in poly$n$ time~\cite{Hal02},
which breaks the Buchman-Williams cryptosystem~\cite{BW88}.

The Abelian Hidden Subgroup Problem~\cite{Mos15} is regarded as the heart of Shor's and related algorithms and yields a superpoly$n$ speedup over the best classical alternative.
Formally, the problem is described by considering a finitely generated Abelian group~$G$
and a subgroup~$H$ with~$G/H$ being finite
along with a function
$f:G\to X$,
for some set~$X$.
Furthermore~$G$ `hides'~$H$;
i.e.,
\begin{equation}
\forall g,g'\in G,
f(g)=f(g')
\Leftrightarrow
gH=g'H.
\end{equation}
The task is to find a set of generators for~$H$
via queries to~$f$;
this is solvable on a quantum computer using $O(\log|G|)$ queries, whereas classically $\Omega(|G|)$ are required~\cite{BL95,NC10},
although solving in one query is possible for some instances~\cite{dBCW02}.

Significant effort has been dedicated to devising quantum algorithms for speeding up solutions to non-Abelian hidden subgroups,
with successful quantum speedup in some cases~\cite{IMS01}
and important implications for problems such as graph isomorphism in the symmetric-group case~\cite{EH99}
or some lattice problems for the dihedral-group 
(symmetry group of the $n$-sided polygon)
case~\cite{Reg02}
with the computational cost for finding a hidden subgroup of the dihedral group reduced to $2^{O(\log\sqrt{n})}$ in the quantum case
compared to the classical cost of $O(\sqrt{n})$~\cite{Kup05}.
This algorithm was further extended to be subexponential in time and polynomial in space~\cite{Reg04}.

Although Shor's algorithm and the variants discussed above can be posed as oracular algorithms,
the analyses above typically consider space and time cost.
On the other hand,
the quantum search algorithm~\cite{Gro97}
is naturally posed as an oracle problem.
This quantum algorithm for unstructured search finds, with high probability,
the unique oracle input whose output is~$1$
(all other $N-1$ oracle outputs are~$0$
although extendable to multiple $1$~\cite{BBHT98}).

First the procedure creates a uniform superposition of states and then applies the quantum oracle 
$\ket{x}\mapsto(-1)^{f(x)}\ket{x}$,
which changes the phase of the marked state.
After sufficiently many iterations, the quantum search algorithm halts and resultant quantum states are measured, 
with measurement of the marked state being highly likely:
A high probability of determining the marked state is achieved with just $O(\sqrt{N})$
queries to the quantum oracle~\cite{BBBV97}
compared to the classical requirement of $O(N)$
queries to the classical oracle~$f$.
The quantum search algorithm is optimal~\cite{BBBV97,Zal99}.

Importantly, although the search-algorithm improvement is only
quadratic  with respect to query complexity, the quadratic improvement is provable,  whereas Shor's algorithm is subexponentially better but based on polynomial hierarchy~\cite{Sto76} arguments rather than on a formal proof.
The quantum search algorithm has amplitude estimation~\cite{BH97,Gro98,BHMT02} at its core and is at the core of quadratically sped-up quantum algorithms such as for enhancing the classical 3-SAT algorithm~\cite{Amb04}.

Another type of task for a quantum computer is to simulate Hamiltonian-generated evolution on a system with~$n$ degrees of freedom
for the Hamiltonian~$H$ promised to have certain properties.
These properties could include being row-computable,
which means that the elements of each row of the Hermitian matrix representing~$H$ are computable, perhaps efficiently.
Another property could be that the Hamiltonian is $k$-local,
which means $H=\sum_{j=1}^mH_j$ 
for each $H_j$ a Hamiltonian acting on up to~$k$ qubits.
Yet another would be that every row of~$H$ contains at most~$d$ nonzero entries,
meaning that it would have to be $d$-sparse.

The problem of quantum simulation, or quantum-state generation~\cite{AT03},
originates with Feynman's motivation for quantum computation~\cite{Fey82}.
One major focus of application is simulating chemical dynamics~\cite{KJL+08}.
Hamiltonian-generated unitary evolution is described by unitary map
$U=\exp(-\text{i}Ht)$
(setting $\hbar=1$).
This evolution is used to map any input state to its $H$-evolved form after time~$t$.
The quantum circuit for simulating this evolution
can be achieved with a poly$(n,t)$-depth quantum circuit~\cite{AT03,Chi04,BACS07,BACS07ch}, hence efficient with respect to these quantities
and not believed to be efficient classically.

Generating the quantum state as a solution is a step towards solving a computational problem but is not the full solution as the end point does not yield  an output string as the solution.
For example, finding the ground-state energy of a $k$-local Hamiltonian is important for physics and chemistry but this problem is hard to solve.
To understand this hardness,
we consider the complexity class \textsf{QMA}.
Roughly speaking \textsf{QMA}
is to \textsf{BQP}
what \textsf{NP} is to \textsf{BPP}
(or to \textsf{P} itself if \textsf{BPP}$=$\textsf{P}).
Solving the ground state of a $k$-local Hamiltonian 
is \textsf{QMA-complete} for $k\geq2$~\cite{KKR04}
(but in~\textsf{P} for $k=1$).
Other applications  such as simulating chemical dynamics~\cite{KJL+08} have been explored.

This complexity result for finding the ground state is analogous to the maximum satisfiability problem (MAX-SAT),
which is an optimisation extension of SAT.
The task for MAX-SAT is to determine the maximum number of clauses of a given Boolean formula expressed in conjunctive normal form that can be made true.
For more than one clause,
MAX-SAT is \textsf{NP-complete}.
For quantum computation,
an extension of MAX-SAT to weighted MAX-SAT~\cite{BF98} is important.
Given a Boolean formula in conjunctive normal form,
with each clause assigned a non-negative weight,
the task is to assign values to the variables that maximize the combined weight of the satisfied clauses.

One application of quantum-state generation has been as a system-of-linear-equations (SLEP ) solver~\cite{HHL09}.
This `HHL', or SLEP, problem~\cite{ANB+22},
accepts an oracle description of $A\in\mathcal{M}_n(\mathbb{R})$
and an efficient description of a vector~$\bm{b}$
(not to be confused with the bit string defined earlier)
and an accuracy parameter~$\varepsilon$.
The task is to compute some property of $f(A)\bm{b}$ for an efficiently computable function~$f$.
Suppose that
$A$ is a polylog$n$-sparse Hermitian matrix.
Furthermore, suppose~$A$ has condition number~$\kappa$
that, generically, quantifies sensitivity of a function's output to variation of input.
Specifically,
for $A\bm{x}=\bm{b}$
(noting that $\|A\|$ is $A$'s largest singular value and, if~$A$ is square,
$\|A^{-1}\|$
is $A$'s smallest singular value),
the ratio of the fractional error of~$\bm{x}$ to the fractional error of~$\bm{b}$
has an upper bound of $\kappa:=\|A\|\cdot\|A^{-1}\|$~\cite{BKW05}.

Under these assumptions,
some expectation values of operators with respect to $f(A)\bm{b}$
can be approximated by quantum computation in $O(\kappa^2\log n)$ time within poly$\left(\nicefrac1\varepsilon\right)$
achieving poly$(\nicefrac1\varepsilon)$ accuracy~\cite{ANB+22}/
The HHL algorithm can be extended to non-sparse matrices but require pre-computation~\cite{KP17,WZP18}.
Practical applications such as for recommendation systems~\cite{KP17}
and principal component analysis~\cite{LMR14}
were later shown to be solvable in poly time by classical randomised algorithms (known as dequantising quantum algorithms) in 2019~\cite{Tan19}
and in 2021~\cite{Tan21}, respectively.
By similar means,
the application of the SLEP method to quantum algorithms for singular-value transformation~\cite{GSLW19,CGJ19}
were also dequantised~\cite{JLS20},
showing the immense challenge in taking the HHL algorithm all the way to practical applications.

Many more quantum algorithms have been constructed and their complexities analysed,
but this discussion conveys some of the key concepts in studies of quantum algorithms and complexity.
Another large area of quantum computation research concerns quantum heuristics for optimisation.
Especially prominent amongst these heuristics is the `Quantum Approximate Optimization Algorithm', or QAOA for short,
applied to approximate solutions of combinatorial optmisiation problems~\cite{FGG14,FGGZ22}.
Early on, this heuristic achieved a better approximation ratio than any known polynomial-time classical algorithm to a to a Bounded Occurrence Constraint Problem~\cite{FGG15}.
Subsequently,
an efficient classical algorithm that
achieves the hardness-of-approximation limit for the approximation ratio was achieved~\cite{BMO+15}.

The quantum-circuit, or gate-based, approach to quantum algorithms is not the only way to perform quantum computation.
Other approaches include measurement-based, or one-way, quantum computation~\cite{RB01},
quantum walks~\cite{CGW13},
and topological quantum computation~\cite{Kit03},
but equivalences between these approaches and the quantum circuit model are known.
One alternative approach,
however,
began with a presumption of inequivalence:
adiabatic quantum computation~\cite{FGGS00,FGG+01}.

Following the adiabatic theorem
(but bearing in mind caveats~\cite{MS04}),
which posits that a system in a ground state remains in that state
%ground state remains in an instantaneous ground state
 if the evolution is sufficiently slow,
adiabatic quantum computation is about commencing with an easy-to-prepare ground state of some Hamiltonian and evolving the system slowly to a new Hamiltonian whose ground state encodes the solution to a problem,
generally one pertinent to combinatorial optimisation.
Adiabatic quantum computation was tested for small instances of an \textsf{NP-complete} problem with indications that adiabatic quantum computation could be advantageous for \textsf{NP-complete} problems~\cite{FGG+01}.

For the minimum eigenvalue gap~$\gamma$
between the ground and first excited states,
the run time of an adiabatic process is $O(\nicefrac1{\gamma^3})$~\cite{JRS07},
which can be improved to~$\tilde{O}(\nicefrac1{\gamma^2})$
if the evolution is sufficiently smooth~\cite{EH12}.
Adiabatic quantum computation is divided into two types:
stochastic vs non-stochastic Hamiltonian,
with the stochastic Hamiltonian being represented by a matrix whose off-diagonal elements in the standard basis are are real and non-positive~\cite{BDOT08}.
Adiabatic quantum computation with non-stochastic Hamiltonians is at least as  
efficient as quantum circuits, whereas restricting to stochastic Hamiltonians is likely
weaker than using quantum circuits~\cite{AvDK+07}
but likely better than classical computation~\cite{Has21}.
\section{Conclusions}
\label{sec:conclusions}

This chapter reviews essential fundamentals of classical computation and quantum mechanics to be able to convey concisely the essence of quantum computation.
The material covered is not exhaustive but rather explains some of the major advances since inception of the field and aims to convey the key concepts.
The quantum algorithm zoo~\cite{QAZ}
is an excellent source of comprehensive information about all quantum algorithms,
and this chapter is partially meant to provide a useful primer for being able to read the material in this zoo.

The industrial side of quantum computation must contend with the parlous state of current quantum computer technologies and so runs whatever quantum algorithms and heuristics are promising and feasible on existing devices. Consequently  much of the material here is beyond the reach of existing commercial quantum computers.
The basics of building a quantum computer are not covered here but can be studied from other sources~\cite{San17}.
Existing quantum computers are said to be following the noisy intermediate-scale quantum paradigm~\cite{Pre18,BCK+22},
which constitutes a valuable step towards large scalable quantum computers of the future.
\bibliography{main}

%apsrev4-2.bst 2019-01-14 (MD) hand-edited version of apsrev4-1.bst
%Control: key (0)
%Control: author (8) initials jnrlst
%Control: editor formatted (1) identically to author
%Control: production of article title (0) allowed
%Control: page (0) single
%Control: year (1) truncated
%Control: production of eprint (0) enabled
\begin{thebibliography}{113}%
\makeatletter
\providecommand \@ifxundefined [1]{%
 \@ifx{#1\undefined}
}%
\providecommand \@ifnum [1]{%
 \ifnum #1\expandafter \@firstoftwo
 \else \expandafter \@secondoftwo
 \fi
}%
\providecommand \@ifx [1]{%
 \ifx #1\expandafter \@firstoftwo
 \else \expandafter \@secondoftwo
 \fi
}%
\providecommand \natexlab [1]{#1}%
\providecommand \enquote  [1]{``#1''}%
\providecommand \bibnamefont  [1]{#1}%
\providecommand \bibfnamefont [1]{#1}%
\providecommand \citenamefont [1]{#1}%
\providecommand \href@noop [0]{\@secondoftwo}%
\providecommand \href [0]{\begingroup \@sanitize@url \@href}%
\providecommand \@href[1]{\@@startlink{#1}\@@href}%
\providecommand \@@href[1]{\endgroup#1\@@endlink}%
\providecommand \@sanitize@url [0]{\catcode `\\12\catcode `\$12\catcode
  `\&12\catcode `\#12\catcode `\^12\catcode `\_12\catcode `\%12\relax}%
\providecommand \@@startlink[1]{}%
\providecommand \@@endlink[0]{}%
\providecommand \url  [0]{\begingroup\@sanitize@url \@url }%
\providecommand \@url [1]{\endgroup\@href {#1}{\urlprefix }}%
\providecommand \urlprefix  [0]{URL }%
\providecommand \Eprint [0]{\href }%
\providecommand \doibase [0]{https://doi.org/}%
\providecommand \selectlanguage [0]{\@gobble}%
\providecommand \bibinfo  [0]{\@secondoftwo}%
\providecommand \bibfield  [0]{\@secondoftwo}%
\providecommand \translation [1]{[#1]}%
\providecommand \BibitemOpen [0]{}%
\providecommand \bibitemStop [0]{}%
\providecommand \bibitemNoStop [0]{.\EOS\space}%
\providecommand \EOS [0]{\spacefactor3000\relax}%
\providecommand \BibitemShut  [1]{\csname bibitem#1\endcsname}%
\let\auto@bib@innerbib\@empty
%</preamble>
\bibitem [{\citenamefont {Sipser}(2012)}]{Sip12}%
  \BibitemOpen
  \bibfield  {author} {\bibinfo {author} {\bibfnamefont {M.}~\bibnamefont
  {Sipser}},\ }\href@noop {} {\emph {\bibinfo {title} {Introduction to the
  {T}heory of {C}omputation}}},\ \bibinfo {edition} {3rd}\ ed.\ (\bibinfo
  {publisher} {Cengage Learning},\ \bibinfo {address} {Boston},\ \bibinfo
  {year} {2012})\BibitemShut {NoStop}%
\bibitem [{\citenamefont {Zhang}\ \emph {et~al.}(2021)\citenamefont {Zhang},
  \citenamefont {Sanders},\ and\ \citenamefont {Sanders}}]{ZSS21}%
  \BibitemOpen
  \bibfield  {author} {\bibinfo {author} {\bibfnamefont {W.-W.}\ \bibnamefont
  {Zhang}}, \bibinfo {author} {\bibfnamefont {Y.~R.}\ \bibnamefont {Sanders}},\
  and\ \bibinfo {author} {\bibfnamefont {B.~C.}\ \bibnamefont {Sanders}},\
  }\bibfield  {title} {\bibinfo {title} {Channel discord and distortion},\
  }\href {https://doi.org/10.1088/1367-2630/ac180a} {\bibfield  {journal}
  {\bibinfo  {journal} {New J. Phys.}\ }\textbf {\bibinfo {volume} {23}},\
  \bibinfo {pages} {083025} (\bibinfo {year} {2021})}\BibitemShut {NoStop}%
\bibitem [{\citenamefont {Blum}\ \emph {et~al.}(1989)\citenamefont {Blum},
  \citenamefont {Shub},\ and\ \citenamefont {Smale}}]{BSS89}%
  \BibitemOpen
  \bibfield  {author} {\bibinfo {author} {\bibfnamefont {L.}~\bibnamefont
  {Blum}}, \bibinfo {author} {\bibfnamefont {M.}~\bibnamefont {Shub}},\ and\
  \bibinfo {author} {\bibfnamefont {S.}~\bibnamefont {Smale}},\ }\bibfield
  {title} {\bibinfo {title} {On a theory of computation and complexity over the
  real numbers: {NP}-completeness, recursive functions and universal
  machines},\ }\href {https://doi.org/10.1090/S0273-0979-1989-15750-9}
  {\bibfield  {journal} {\bibinfo  {journal} {Bull. Amer. Math. Soc.}\ }\textbf
  {\bibinfo {volume} {21}},\ \bibinfo {pages} {1} (\bibinfo {year}
  {1989})}\BibitemShut {NoStop}%
\bibitem [{\citenamefont {Turing}(2004)}]{Tur36}%
  \BibitemOpen
  \bibfield  {author} {\bibinfo {author} {\bibfnamefont {A.}~\bibnamefont
  {Turing}},\ }\bibfield  {title} {\bibinfo {title} {{58On Computable Numbers,
  with an Application to the Entscheidungsproblem (1936)}},\ }in\ \href
  {https://doi.org/10.1093/oso/9780198250791.003.0005} {\emph {\bibinfo
  {booktitle} {{The Essential Turing}}}}\ (\bibinfo  {publisher} {Oxford
  University Press},\ \bibinfo {year} {2004})\BibitemShut {NoStop}%
\bibitem [{\citenamefont {Thomson}(1876)}]{Tho76}%
  \BibitemOpen
  \bibfield  {author} {\bibinfo {author} {\bibfnamefont {J.}~\bibnamefont
  {Thomson}},\ }\bibfield  {title} {\bibinfo {title} {Mechanical integration of
  linear differential equations of the second order with variable
  coefficients},\ }\href {https://doi.org/10.1098/rspl.1875.0035. S2CID
  62694536} {\bibfield  {journal} {\bibinfo  {journal} {Proc. R. Soc. Lond.}\
  }\textbf {\bibinfo {volume} {XXIV}},\ \bibinfo {pages} {262} (\bibinfo {year}
  {1876})}\BibitemShut {NoStop}%
\bibitem [{\citenamefont {Rad\'{o}}(1962)}]{R62}%
  \BibitemOpen
  \bibfield  {author} {\bibinfo {author} {\bibfnamefont {T.}~\bibnamefont
  {Rad\'{o}}},\ }\bibfield  {title} {\bibinfo {title} {On non-computable
  functions},\ }\href {https://doi.org/10.1002/j.1538-7305.1962.tb00480.x}
  {\bibfield  {journal} {\bibinfo  {journal} {Bell Labs Tech. J.}\ }\textbf
  {\bibinfo {volume} {41}},\ \bibinfo {pages} {877} (\bibinfo {year}
  {1962})}\BibitemShut {NoStop}%
\bibitem [{\citenamefont {Goldreich}(2008)}]{Gol08}%
  \BibitemOpen
  \bibfield  {author} {\bibinfo {author} {\bibfnamefont {O.}~\bibnamefont
  {Goldreich}},\ }\href@noop {} {\emph {\bibinfo {title} {Computational
  Complexity: A Conceptual Perspective}}}\ (\bibinfo  {publisher} {Cambridge
  University Press},\ \bibinfo {address} {New York},\ \bibinfo {year}
  {2008})\BibitemShut {NoStop}%
\bibitem [{\citenamefont {Landauer}(1961)}]{Lan61}%
  \BibitemOpen
  \bibfield  {author} {\bibinfo {author} {\bibfnamefont {R.}~\bibnamefont
  {Landauer}},\ }\bibfield  {title} {\bibinfo {title} {Irreversibility and heat
  generation in the computing process},\ }\href
  {https://doi.org/10.1147/rd.53.0183} {\bibfield  {journal} {\bibinfo
  {journal} {IBM J. Res. Dev.}\ }\textbf {\bibinfo {volume} {5}},\ \bibinfo
  {pages} {183} (\bibinfo {year} {1961})}\BibitemShut {NoStop}%
\bibitem [{\citenamefont {Bennett}(1973)}]{Ben73}%
  \BibitemOpen
  \bibfield  {author} {\bibinfo {author} {\bibfnamefont {C.~H.}\ \bibnamefont
  {Bennett}},\ }\bibfield  {title} {\bibinfo {title} {Logical reversibility of
  computation},\ }\href {https://doi.org/10.1147/rd.176.0525} {\bibfield
  {journal} {\bibinfo  {journal} {IBM J. Res. Dev.}\ }\textbf {\bibinfo
  {volume} {17}},\ \bibinfo {pages} {525} (\bibinfo {year} {1973})}\BibitemShut
  {NoStop}%
\bibitem [{\citenamefont {Deutsch}(1985)}]{Deu85}%
  \BibitemOpen
  \bibfield  {author} {\bibinfo {author} {\bibfnamefont {D.}~\bibnamefont
  {Deutsch}},\ }\bibfield  {title} {\bibinfo {title} {Quantum theory, the
  {C}hurch-{T}uring principle and the universal quantum computer},\ }\href@noop
  {} {\bibfield  {journal} {\bibinfo  {journal} {Proc. R. Soc. Lond. A}\
  }\textbf {\bibinfo {volume} {400}},\ \bibinfo {pages} {97} (\bibinfo {year}
  {1985})}\BibitemShut {NoStop}%
\bibitem [{\citenamefont {Kleene}(1952)}]{Kle52}%
  \BibitemOpen
  \bibfield  {author} {\bibinfo {author} {\bibfnamefont {S.~C.}\ \bibnamefont
  {Kleene}},\ }\href@noop {} {\emph {\bibinfo {title} {Introduction to
  {M}etamathematics}}},\ University series in higher mathematics\ (\bibinfo
  {publisher} {van Nostrand},\ \bibinfo {year} {1952})\BibitemShut {NoStop}%
\bibitem [{\citenamefont {Bang-Jensen}\ and\ \citenamefont
  {Gutin}(2009)}]{BG09}%
  \BibitemOpen
  \bibfield  {author} {\bibinfo {author} {\bibfnamefont {J.}~\bibnamefont
  {Bang-Jensen}}\ and\ \bibinfo {author} {\bibfnamefont {G.~Z.}\ \bibnamefont
  {Gutin}},\ }\bibinfo {title} {Classes of digraphs},\ in\ \href
  {https://doi.org/10.1007/978-1-84800-998-1_2} {\emph {\bibinfo {booktitle}
  {Digraphs: Theory, Algorithms and Applications}}}\ (\bibinfo  {publisher}
  {Springer London},\ \bibinfo {address} {London},\ \bibinfo {year} {2009})\
  pp.\ \bibinfo {pages} {31--86}\BibitemShut {NoStop}%
\bibitem [{\citenamefont {Pippenger}\ and\ \citenamefont
  {Fischer}(1979)}]{PF79}%
  \BibitemOpen
  \bibfield  {author} {\bibinfo {author} {\bibfnamefont {N.}~\bibnamefont
  {Pippenger}}\ and\ \bibinfo {author} {\bibfnamefont {M.~J.}\ \bibnamefont
  {Fischer}},\ }\bibfield  {title} {\bibinfo {title} {Relations among
  complexity measures},\ }\href {https://doi.org/10.1145/322123.322138}
  {\bibfield  {journal} {\bibinfo  {journal} {J. ACM}\ }\textbf {\bibinfo
  {volume} {26}},\ \bibinfo {pages} {361} (\bibinfo {year} {1979})}\BibitemShut
  {NoStop}%
\bibitem [{\citenamefont {Cormen}\ \emph {et~al.}(2009)\citenamefont {Cormen},
  \citenamefont {Leiserson}, \citenamefont {Rivest},\ and\ \citenamefont
  {Stein}}]{CLRS09}%
  \BibitemOpen
  \bibfield  {author} {\bibinfo {author} {\bibfnamefont {T.~H.}\ \bibnamefont
  {Cormen}}, \bibinfo {author} {\bibfnamefont {C.~E.}\ \bibnamefont
  {Leiserson}}, \bibinfo {author} {\bibfnamefont {R.~L.}\ \bibnamefont
  {Rivest}},\ and\ \bibinfo {author} {\bibfnamefont {C.}~\bibnamefont
  {Stein}},\ }\href@noop {} {\emph {\bibinfo {title} {Introduction to
  Algorithms}}},\ \bibinfo {edition} {3rd}\ ed.\ (\bibinfo  {publisher} {MIT
  Press},\ \bibinfo {year} {2009})\BibitemShut {NoStop}%
\bibitem [{\citenamefont {Knuth}(1976)}]{Knu76}%
  \BibitemOpen
  \bibfield  {author} {\bibinfo {author} {\bibfnamefont {D.~E.}\ \bibnamefont
  {Knuth}},\ }\bibfield  {title} {\bibinfo {title} {Big omicron and big omega
  and big theta},\ }\href {https://doi.org/10.1145/1008328.1008329} {\bibfield
  {journal} {\bibinfo  {journal} {SIGACT News}\ }\textbf {\bibinfo {volume}
  {8}},\ \bibinfo {pages} {18} (\bibinfo {year} {1976})}\BibitemShut {NoStop}%
\bibitem [{\citenamefont {Cobham}(1965)}]{Cob65}%
  \BibitemOpen
  \bibfield  {author} {\bibinfo {author} {\bibfnamefont {A.}~\bibnamefont
  {Cobham}},\ }\bibfield  {title} {\bibinfo {title} {The intrinsic
  computational difficulty of functions},\ }in\ \href@noop {} {\emph {\bibinfo
  {booktitle} {Logic, methodology and philosophy of science, Proc.\ 1964
  International Congress}}},\ \bibinfo {series and number} {Studies in logic
  and the foundations of mathematics},\ \bibinfo {editor} {edited by\ \bibinfo
  {editor} {\bibfnamefont {Y.}~\bibnamefont {Bar-Hillel}}}\ (\bibinfo
  {publisher} {North-Holland},\ \bibinfo {address} {Amsterdam},\ \bibinfo
  {year} {1965})\ pp.\ \bibinfo {pages} {24--30}\BibitemShut {NoStop}%
\bibitem [{\citenamefont {Stockmeyer}(1976)}]{Sto76}%
  \BibitemOpen
  \bibfield  {author} {\bibinfo {author} {\bibfnamefont {L.~J.}\ \bibnamefont
  {Stockmeyer}},\ }\bibfield  {title} {\bibinfo {title} {The polynomial-time
  hierarchy},\ }\href {https://doi.org/10.1016/0304-3975(76)90061-X} {\bibfield
   {journal} {\bibinfo  {journal} {Theor. Comput. Sci.}\ }\textbf {\bibinfo
  {volume} {3}},\ \bibinfo {pages} {1} (\bibinfo {year} {1976})}\BibitemShut
  {NoStop}%
\bibitem [{\citenamefont {Fortnow}(2009)}]{For09}%
  \BibitemOpen
  \bibfield  {author} {\bibinfo {author} {\bibfnamefont {L.}~\bibnamefont
  {Fortnow}},\ }\bibfield  {title} {\bibinfo {title} {The status of the {P}
  versus {NP} problem},\ }\href {https://doi.org/10.1145/1562164.1562186}
  {\bibfield  {journal} {\bibinfo  {journal} {Commun. ACM}\ }\textbf {\bibinfo
  {volume} {52}},\ \bibinfo {pages} {78} (\bibinfo {year} {2009})}\BibitemShut
  {NoStop}%
\bibitem [{\citenamefont {Lawler}\ \emph {et~al.}(1991)\citenamefont {Lawler},
  \citenamefont {Lenstra}, \citenamefont {Rinnooy~Kan},\ and\ \citenamefont
  {Shmoys}}]{LLRS91}%
  \BibitemOpen
  \bibfield  {author} {\bibinfo {author} {\bibfnamefont {E.~L.}\ \bibnamefont
  {Lawler}}, \bibinfo {author} {\bibfnamefont {J.~K.}\ \bibnamefont {Lenstra}},
  \bibinfo {author} {\bibfnamefont {A.~H.~G.}\ \bibnamefont {Rinnooy~Kan}},\
  and\ \bibinfo {author} {\bibfnamefont {D.~B.}\ \bibnamefont {Shmoys}},\
  }\href@noop {} {\emph {\bibinfo {title} {The Traveling Salesman Problem: A
  Guided Tour of Combinatorial Optimization}}}\ (\bibinfo  {publisher}
  {Wiley},\ \bibinfo {year} {1991})\BibitemShut {NoStop}%
\bibitem [{\citenamefont {Sahni}\ and\ \citenamefont {Gonzalez}(1976)}]{SG76}%
  \BibitemOpen
  \bibfield  {author} {\bibinfo {author} {\bibfnamefont {S.}~\bibnamefont
  {Sahni}}\ and\ \bibinfo {author} {\bibfnamefont {T.}~\bibnamefont
  {Gonzalez}},\ }\bibfield  {title} {\bibinfo {title} {P-complete approximation
  problems},\ }\href {https://doi.org/10.1145/321958.321975} {\bibfield
  {journal} {\bibinfo  {journal} {J. ACM}\ }\textbf {\bibinfo {volume} {23}},\
  \bibinfo {pages} {555} (\bibinfo {year} {1976})}\BibitemShut {NoStop}%
\bibitem [{\citenamefont {Gill}(1977)}]{Gil77}%
  \BibitemOpen
  \bibfield  {author} {\bibinfo {author} {\bibfnamefont {J.}~\bibnamefont
  {Gill}},\ }\bibfield  {title} {\bibinfo {title} {Computational complexity of
  probabilistic turing machines},\ }\href@noop {} {\bibfield  {journal}
  {\bibinfo  {journal} {SIAM J. Comput.}\ }\textbf {\bibinfo {volume} {6}},\
  \bibinfo {pages} {675} (\bibinfo {year} {1977})}\BibitemShut {NoStop}%
\bibitem [{\citenamefont {Vajirani}(2001)}]{Vaz01}%
  \BibitemOpen
  \bibfield  {author} {\bibinfo {author} {\bibfnamefont {V.~V.}\ \bibnamefont
  {Vajirani}},\ }\href@noop {} {\emph {\bibinfo {title} {Approximation
  Algorithms}}}\ (\bibinfo  {publisher} {Springer-Verlag},\ \bibinfo {address}
  {Berlin},\ \bibinfo {year} {2001})\BibitemShut {NoStop}%
\bibitem [{\citenamefont {Bravyi}(2011)}]{Bra11}%
  \BibitemOpen
  \bibfield  {author} {\bibinfo {author} {\bibfnamefont {S.}~\bibnamefont
  {Bravyi}},\ }\bibinfo {title} {Cross {D}isciplinary {A}dvances in {Q}uantum
  {C}omputing}\ (\bibinfo  {publisher} {American Mathematical Society},\
  \bibinfo {year} {2011})\ Chap.\ \bibinfo {chapter} {Efficient algorithm for a
  quantum analogue of 2-SAT}\BibitemShut {NoStop}%
\bibitem [{\citenamefont {Krom}(1967)}]{Kro67}%
  \BibitemOpen
  \bibfield  {author} {\bibinfo {author} {\bibfnamefont {M.~R.}\ \bibnamefont
  {Krom}},\ }\bibfield  {title} {\bibinfo {title} {The decision problem for a
  class of first-order formulas in which all disjunctions are binary},\ }\href
  {https://doi.org/10.1002/malq.19670130104} {\bibfield  {journal} {\bibinfo
  {journal} {Math. Log. Q.}\ }\textbf {\bibinfo {volume} {13}},\ \bibinfo
  {pages} {15} (\bibinfo {year} {1967})}\BibitemShut {NoStop}%
\bibitem [{\citenamefont {Garey}\ and\ \citenamefont {Johnson}(1979)}]{GJ79}%
  \BibitemOpen
  \bibfield  {author} {\bibinfo {author} {\bibfnamefont {M.~R.}\ \bibnamefont
  {Garey}}\ and\ \bibinfo {author} {\bibfnamefont {D.~S.}\ \bibnamefont
  {Johnson}},\ }\href@noop {} {\emph {\bibinfo {title} {Computers and
  {I}ntractability: {A} {G}uide to the {T}heory of {NP}-completeness}}}\
  (\bibinfo  {publisher} {Freeman},\ \bibinfo {address} {San Francisco},\
  \bibinfo {year} {1979})\BibitemShut {NoStop}%
\bibitem [{\citenamefont {Bennett}(2000)}]{Ben00}%
  \BibitemOpen
  \bibfield  {author} {\bibinfo {author} {\bibfnamefont {C.~H.}\ \bibnamefont
  {Bennett}},\ }\bibfield  {title} {\bibinfo {title} {Notes on the history of
  reversible computation},\ }\href {https://doi.org/10.1147/rd.441.0270}
  {\bibfield  {journal} {\bibinfo  {journal} {IBM J. Res. Dev.}\ }\textbf
  {\bibinfo {volume} {44}},\ \bibinfo {pages} {270} (\bibinfo {year}
  {2000})}\BibitemShut {NoStop}%
\bibitem [{\citenamefont {Toffoli}(1980)}]{Tof80}%
  \BibitemOpen
  \bibfield  {author} {\bibinfo {author} {\bibfnamefont {T.}~\bibnamefont
  {Toffoli}},\ }\bibfield  {title} {\bibinfo {title} {Reversible computing},\
  }in\ \href@noop {} {\emph {\bibinfo {booktitle} {Automata, Languages and
  Programming}}},\ \bibinfo {editor} {edited by\ \bibinfo {editor}
  {\bibfnamefont {J.}~\bibnamefont {de~Bakker}}\ and\ \bibinfo {editor}
  {\bibfnamefont {J.}~\bibnamefont {van Leeuwen}}}\ (\bibinfo  {publisher}
  {Springer},\ \bibinfo {address} {Berlin},\ \bibinfo {year} {1980})\ pp.\
  \bibinfo {pages} {632--644}\BibitemShut {NoStop}%
\bibitem [{\citenamefont {Perumalla}(2013)}]{Per13}%
  \BibitemOpen
  \bibfield  {author} {\bibinfo {author} {\bibfnamefont {K.~S.}\ \bibnamefont
  {Perumalla}},\ }\href@noop {} {\emph {\bibinfo {title} {Introduction to
  Reversible Computing}}},\ Computational Science\ (\bibinfo  {publisher}
  {Chapman \&\ Hall/CRC},\ \bibinfo {year} {2013})\BibitemShut {NoStop}%
\bibitem [{\citenamefont {Maruyama}\ \emph {et~al.}(2009)\citenamefont
  {Maruyama}, \citenamefont {Nori},\ and\ \citenamefont {Vedral}}]{MNV09}%
  \BibitemOpen
  \bibfield  {author} {\bibinfo {author} {\bibfnamefont {K.}~\bibnamefont
  {Maruyama}}, \bibinfo {author} {\bibfnamefont {F.}~\bibnamefont {Nori}},\
  and\ \bibinfo {author} {\bibfnamefont {V.}~\bibnamefont {Vedral}},\
  }\bibfield  {title} {\bibinfo {title} {Colloquium: The physics of {M}axwell's
  demon and information},\ }\href {https://doi.org/10.1103/RevModPhys.81.1}
  {\bibfield  {journal} {\bibinfo  {journal} {Rev. Mod. Phys.}\ }\textbf
  {\bibinfo {volume} {81}},\ \bibinfo {pages} {1} (\bibinfo {year}
  {2009})}\BibitemShut {NoStop}%
\bibitem [{\citenamefont {Ray}\ and\ \citenamefont {Crutchfield}(2020)}]{RC20}%
  \BibitemOpen
  \bibfield  {author} {\bibinfo {author} {\bibfnamefont {K.~J.}\ \bibnamefont
  {Ray}}\ and\ \bibinfo {author} {\bibfnamefont {J.~P.}\ \bibnamefont
  {Crutchfield}},\ }\bibfield  {title} {\bibinfo {title} {Variations on a
  demonic theme: Szilard’s other engines},\ }\href
  {https://doi.org/10.1063/5.0012052} {\bibfield  {journal} {\bibinfo
  {journal} {Chaos}\ }\textbf {\bibinfo {volume} {30}},\ \bibinfo {pages}
  {093105} (\bibinfo {year} {2020})}\BibitemShut {NoStop}%
\bibitem [{\citenamefont {Carette}\ \emph {et~al.}(2022)\citenamefont
  {Carette}, \citenamefont {James},\ and\ \citenamefont {Sabry}}]{CJS22}%
  \BibitemOpen
  \bibfield  {author} {\bibinfo {author} {\bibfnamefont {J.}~\bibnamefont
  {Carette}}, \bibinfo {author} {\bibfnamefont {R.~P.}\ \bibnamefont {James}},\
  and\ \bibinfo {author} {\bibfnamefont {A.}~\bibnamefont {Sabry}},\ }\bibfield
   {title} {\bibinfo {title} {Chapter two - embracing the laws of physics:
  Three reversible models of computation}\ }(\bibinfo  {publisher} {Elsevier},\
  \bibinfo {year} {2022})\ pp.\ \bibinfo {pages} {15--63}\BibitemShut {NoStop}%
\bibitem [{\citenamefont {Fredkin}\ and\ \citenamefont {Toffoli}(1982)}]{FT82}%
  \BibitemOpen
  \bibfield  {author} {\bibinfo {author} {\bibfnamefont {E.}~\bibnamefont
  {Fredkin}}\ and\ \bibinfo {author} {\bibfnamefont {T.}~\bibnamefont
  {Toffoli}},\ }\bibfield  {title} {\bibinfo {title} {Conservative logic},\
  }\href {https://doi.org/10.1007/BF01857727} {\bibfield  {journal} {\bibinfo
  {journal} {Int. J. Theor. Phys.}\ }\textbf {\bibinfo {volume} {21}},\
  \bibinfo {pages} {219} (\bibinfo {year} {1982})}\BibitemShut {NoStop}%
\bibitem [{\citenamefont {Sasao}\ and\ \citenamefont {Kinoshita}(1979)}]{SK79}%
  \BibitemOpen
  \bibfield  {author} {\bibinfo {author} {\bibfnamefont {T.}~\bibnamefont
  {Sasao}}\ and\ \bibinfo {author} {\bibfnamefont {K.}~\bibnamefont
  {Kinoshita}},\ }\bibfield  {title} {\bibinfo {title} {Conservative logic
  elements and their universality},\ }\href@noop {} {\bibfield  {journal}
  {\bibinfo  {journal} {IEEE Trans. Comput.}\ } (\bibinfo {year}
  {1979})}\BibitemShut {NoStop}%
\bibitem [{\citenamefont {Strocchi}(2008)}]{Str08}%
  \BibitemOpen
  \bibfield  {author} {\bibinfo {author} {\bibfnamefont {F.}~\bibnamefont
  {Strocchi}},\ }\href@noop {} {\emph {\bibinfo {title} {{An Introduction to
  the Mathematical Structure of Quantum Mechanics: A Short Course for
  Mathematicians}}}},\ \bibinfo {edition} {2nd}\ ed.,\ Advanced Series in
  Mathematical Physics\ (\bibinfo  {publisher} {World Scientific},\ \bibinfo
  {address} {Singapore},\ \bibinfo {year} {2008})\BibitemShut {NoStop}%
\bibitem [{\citenamefont {Gelfand}\ and\ \citenamefont {Neumark}(1943)}]{GN43}%
  \BibitemOpen
  \bibfield  {author} {\bibinfo {author} {\bibfnamefont {I.}~\bibnamefont
  {Gelfand}}\ and\ \bibinfo {author} {\bibfnamefont {M.}~\bibnamefont
  {Neumark}},\ }\bibfield  {title} {\bibinfo {title} {On the imbedding of
  normed rings into the ring of operators in {Hilbert} space},\ }\href@noop {}
  {\bibfield  {journal} {\bibinfo  {journal} {Rec. Math. [Mat. Sbornik] N.S.}\
  }\textbf {\bibinfo {volume} {12}},\ \bibinfo {pages} {197} (\bibinfo {year}
  {1943})}\BibitemShut {NoStop}%
\bibitem [{\citenamefont {Segal}(1947)}]{Seg47}%
  \BibitemOpen
  \bibfield  {author} {\bibinfo {author} {\bibfnamefont {I.~E.}\ \bibnamefont
  {Segal}},\ }\bibfield  {title} {\bibinfo {title} {Irreducible representations
  of operator algebras},\ }\href@noop {} {\bibfield  {journal} {\bibinfo
  {journal} {Bull. Amer. Math. Soc.}\ }\textbf {\bibinfo {volume} {53}},\
  \bibinfo {pages} {73} (\bibinfo {year} {1947})}\BibitemShut {NoStop}%
\bibitem [{\citenamefont {Schumacher}(1995)}]{Sch95}%
  \BibitemOpen
  \bibfield  {author} {\bibinfo {author} {\bibfnamefont {B.}~\bibnamefont
  {Schumacher}},\ }\bibfield  {title} {\bibinfo {title} {Quantum coding},\
  }\href {https://doi.org/10.1103/PhysRevA.51.2738} {\bibfield  {journal}
  {\bibinfo  {journal} {Phys. Rev. A}\ }\textbf {\bibinfo {volume} {51}},\
  \bibinfo {pages} {2738} (\bibinfo {year} {1995})}\BibitemShut {NoStop}%
\bibitem [{\citenamefont {Brody}\ and\ \citenamefont {Hughston}(2001)}]{BH01}%
  \BibitemOpen
  \bibfield  {author} {\bibinfo {author} {\bibfnamefont {D.~C.}\ \bibnamefont
  {Brody}}\ and\ \bibinfo {author} {\bibfnamefont {L.~P.}\ \bibnamefont
  {Hughston}},\ }\bibfield  {title} {\bibinfo {title} {Geometric quantum
  mechanics},\ }\href {https://doi.org/10.1016/S0393-0440(00)00052-8}
  {\bibfield  {journal} {\bibinfo  {journal} {J. Geom. Phys.}\ }\textbf
  {\bibinfo {volume} {38}},\ \bibinfo {pages} {19} (\bibinfo {year}
  {2001})}\BibitemShut {NoStop}%
\bibitem [{\citenamefont {Watrous}(2018)}]{Wat18}%
  \BibitemOpen
  \bibfield  {author} {\bibinfo {author} {\bibfnamefont {J.}~\bibnamefont
  {Watrous}},\ }\href {https://doi.org/10.1017/9781316848142} {\emph {\bibinfo
  {title} {The Theory of Quantum Information}}}\ (\bibinfo  {publisher}
  {Cambridge University Press},\ \bibinfo {address} {Cambridge},\ \bibinfo
  {year} {2018})\BibitemShut {NoStop}%
\bibitem [{\citenamefont {Dawson}\ and\ \citenamefont {Nielsen}(2006)}]{DN06}%
  \BibitemOpen
  \bibfield  {author} {\bibinfo {author} {\bibfnamefont {C.~M.}\ \bibnamefont
  {Dawson}}\ and\ \bibinfo {author} {\bibfnamefont {M.~A.}\ \bibnamefont
  {Nielsen}},\ }\bibfield  {title} {\bibinfo {title} {The {S}olovay-{K}itaev
  algorithm},\ }\href@noop {} {\bibfield  {journal} {\bibinfo  {journal}
  {Quantum Inf. Comput.}\ }\textbf {\bibinfo {volume} {6}},\ \bibinfo {pages}
  {81} (\bibinfo {year} {2006})}\BibitemShut {NoStop}%
\bibitem [{\citenamefont {Barenco}\ \emph {et~al.}(1995)\citenamefont
  {Barenco}, \citenamefont {Bennett}, \citenamefont {Cleve}, \citenamefont
  {DiVincenzo}, \citenamefont {Margolus}, \citenamefont {Shor}, \citenamefont
  {Sleator}, \citenamefont {Smolin},\ and\ \citenamefont
  {Weinfurter}}]{BBC+95}%
  \BibitemOpen
  \bibfield  {author} {\bibinfo {author} {\bibfnamefont {A.}~\bibnamefont
  {Barenco}}, \bibinfo {author} {\bibfnamefont {C.~H.}\ \bibnamefont
  {Bennett}}, \bibinfo {author} {\bibfnamefont {R.}~\bibnamefont {Cleve}},
  \bibinfo {author} {\bibfnamefont {D.~P.}\ \bibnamefont {DiVincenzo}},
  \bibinfo {author} {\bibfnamefont {N.}~\bibnamefont {Margolus}}, \bibinfo
  {author} {\bibfnamefont {P.}~\bibnamefont {Shor}}, \bibinfo {author}
  {\bibfnamefont {T.}~\bibnamefont {Sleator}}, \bibinfo {author} {\bibfnamefont
  {J.~A.}\ \bibnamefont {Smolin}},\ and\ \bibinfo {author} {\bibfnamefont
  {H.}~\bibnamefont {Weinfurter}},\ }\bibfield  {title} {\bibinfo {title}
  {Elementary gates for quantum computation},\ }\href
  {https://doi.org/10.1103/PhysRevA.52.3457} {\bibfield  {journal} {\bibinfo
  {journal} {Phys. Rev. A}\ }\textbf {\bibinfo {volume} {52}},\ \bibinfo
  {pages} {3457} (\bibinfo {year} {1995})}\BibitemShut {NoStop}%
\bibitem [{\citenamefont {Mosca}(2009)}]{Mos09}%
  \BibitemOpen
  \bibfield  {author} {\bibinfo {author} {\bibfnamefont {M.}~\bibnamefont
  {Mosca}},\ }\bibinfo {title} {Quantum algorithms},\ in\ \href
  {https://doi.org/10.1007/978-0-387-30440-3_423} {\emph {\bibinfo {booktitle}
  {Encyclopedia of Complexity and Systems Science}}},\ \bibinfo {editor}
  {edited by\ \bibinfo {editor} {\bibfnamefont {R.~A.}\ \bibnamefont
  {Meyers}}}\ (\bibinfo  {publisher} {Springer},\ \bibinfo {address} {New
  York},\ \bibinfo {year} {2009})\ pp.\ \bibinfo {pages}
  {7088--7118}\BibitemShut {NoStop}%
\bibitem [{\citenamefont {Bianchi}\ \emph {et~al.}(2009)\citenamefont
  {Bianchi}, \citenamefont {Dorigo}, \citenamefont {Gambardella},\ and\
  \citenamefont {Gutjahr}}]{BDGG09}%
  \BibitemOpen
  \bibfield  {author} {\bibinfo {author} {\bibfnamefont {L.}~\bibnamefont
  {Bianchi}}, \bibinfo {author} {\bibfnamefont {M.}~\bibnamefont {Dorigo}},
  \bibinfo {author} {\bibfnamefont {L.~M.}\ \bibnamefont {Gambardella}},\ and\
  \bibinfo {author} {\bibfnamefont {W.~J.}\ \bibnamefont {Gutjahr}},\
  }\bibfield  {title} {\bibinfo {title} {A survey on metaheuristics for
  stochastic combinatorial optimization},\ }\href
  {https://doi.org/10.1007/s11047-008-9098-4} {\bibfield  {journal} {\bibinfo
  {journal} {Nat. Comput.}\ }\textbf {\bibinfo {volume} {8}},\ \bibinfo {pages}
  {239} (\bibinfo {year} {2009})}\BibitemShut {NoStop}%
\bibitem [{\citenamefont {Montanaro}(2016)}]{Mon16}%
  \BibitemOpen
  \bibfield  {author} {\bibinfo {author} {\bibfnamefont {A.}~\bibnamefont
  {Montanaro}},\ }\bibfield  {title} {\bibinfo {title} {Quantum algorithms: an
  overview},\ }\href {https://doi.org/10.1038/npjqi.2015.23} {\bibfield
  {journal} {\bibinfo  {journal} {npj Quantum Inf.}\ }\textbf {\bibinfo
  {volume} {2}},\ \bibinfo {pages} {15023} (\bibinfo {year}
  {2016})}\BibitemShut {NoStop}%
\bibitem [{\citenamefont {Jordan}(2023)}]{QAZ}%
  \BibitemOpen
  \bibfield  {author} {\bibinfo {author} {\bibfnamefont {S.}~\bibnamefont
  {Jordan}},\ }\href {math.nist.gov/quantum/zoo/} {\bibinfo {title} {The
  quantum algorithm zoo}} (\bibinfo {year} {2023})\BibitemShut {NoStop}%
\bibitem [{\citenamefont {Simon}(1997)}]{Sim97}%
  \BibitemOpen
  \bibfield  {author} {\bibinfo {author} {\bibfnamefont {D.}~\bibnamefont
  {Simon}},\ }\bibfield  {title} {\bibinfo {title} {On the power of quantum
  computation},\ }\href@noop {} {\bibfield  {journal} {\bibinfo  {journal}
  {SIAM J. Comput.}\ }\textbf {\bibinfo {volume} {26}},\ \bibinfo {pages}
  {1474} (\bibinfo {year} {1997})}\BibitemShut {NoStop}%
\bibitem [{\citenamefont {Bernstein}\ and\ \citenamefont
  {Vazirani}(1997)}]{BV97}%
  \BibitemOpen
  \bibfield  {author} {\bibinfo {author} {\bibfnamefont {E.}~\bibnamefont
  {Bernstein}}\ and\ \bibinfo {author} {\bibfnamefont {U.}~\bibnamefont
  {Vazirani}},\ }\bibfield  {title} {\bibinfo {title} {Quantum complexity
  theory},\ }\href@noop {} {\bibfield  {journal} {\bibinfo  {journal} {SIAM J.
  Comput.}\ }\textbf {\bibinfo {volume} {26}},\ \bibinfo {pages} {1411}
  (\bibinfo {year} {1997})}\BibitemShut {NoStop}%
\bibitem [{\citenamefont {Deutsch}\ and\ \citenamefont {Jozsa}(1992)}]{DJ92}%
  \BibitemOpen
  \bibfield  {author} {\bibinfo {author} {\bibfnamefont {D.}~\bibnamefont
  {Deutsch}}\ and\ \bibinfo {author} {\bibfnamefont {R.}~\bibnamefont
  {Jozsa}},\ }\bibfield  {title} {\bibinfo {title} {Rapid solution of problems
  by quantum computation},\ }\href {https://doi.org/10.1098/rspa.1992.0167}
  {\bibfield  {journal} {\bibinfo  {journal} {Proc. R. Soc. Lond. A}\ }\textbf
  {\bibinfo {volume} {439}},\ \bibinfo {pages} {553} (\bibinfo {year}
  {1992})}\BibitemShut {NoStop}%
\bibitem [{\citenamefont {Shor}(1994)}]{Sho94}%
  \BibitemOpen
  \bibfield  {author} {\bibinfo {author} {\bibfnamefont {P.~W.}\ \bibnamefont
  {Shor}},\ }\bibfield  {title} {\bibinfo {title} {Algorithms for quantum
  computation: Discrete logarithms and factoring},\ }in\ \href@noop {} {\emph
  {\bibinfo {booktitle} {Proc.\ 35th Annual Symposium on Foundations of
  Computer Science, SFCS `94'}}}\ (\bibinfo {year} {1994})\ pp.\ \bibinfo
  {pages} {124--134}\BibitemShut {NoStop}%
\bibitem [{\citenamefont {Shor}(1997)}]{Sho97}%
  \BibitemOpen
  \bibfield  {author} {\bibinfo {author} {\bibfnamefont {P.~W.}\ \bibnamefont
  {Shor}},\ }\bibfield  {title} {\bibinfo {title} {Polynomial-time algorithms
  for prime factorization and discrete logarithms on a quantum computer},\
  }\href {https://doi.org/10.1137/S0097539795293172} {\bibfield  {journal}
  {\bibinfo  {journal} {SIAM J. Sci. Comput.}\ }\textbf {\bibinfo {volume}
  {26}},\ \bibinfo {pages} {1484} (\bibinfo {year} {1997})}\BibitemShut
  {NoStop}%
\bibitem [{\citenamefont {Beckman}\ \emph {et~al.}(1996)\citenamefont
  {Beckman}, \citenamefont {Chari}, \citenamefont {Devabhaktuni},\ and\
  \citenamefont {Preskill}}]{BCDP96}%
  \BibitemOpen
  \bibfield  {author} {\bibinfo {author} {\bibfnamefont {D.}~\bibnamefont
  {Beckman}}, \bibinfo {author} {\bibfnamefont {A.~N.}\ \bibnamefont {Chari}},
  \bibinfo {author} {\bibfnamefont {S.}~\bibnamefont {Devabhaktuni}},\ and\
  \bibinfo {author} {\bibfnamefont {J.}~\bibnamefont {Preskill}},\ }\bibfield
  {title} {\bibinfo {title} {Efficient networks for quantum factoring},\ }\href
  {https://doi.org/10.1103/PhysRevA.54.1034} {\bibfield  {journal} {\bibinfo
  {journal} {Phys. Rev. A}\ }\textbf {\bibinfo {volume} {54}},\ \bibinfo
  {pages} {1034} (\bibinfo {year} {1996})}\BibitemShut {NoStop}%
\bibitem [{\citenamefont {Pomerance}(1996)}]{Pom96}%
  \BibitemOpen
  \bibfield  {author} {\bibinfo {author} {\bibfnamefont {C.}~\bibnamefont
  {Pomerance}},\ }\bibfield  {title} {\bibinfo {title} {A tale of two sieves},\
  }\href@noop {} {\bibfield  {journal} {\bibinfo  {journal} {Not. Am. Math.
  Soc.}\ }\textbf {\bibinfo {volume} {43}},\ \bibinfo {pages} {1473} (\bibinfo
  {year} {1996})}\BibitemShut {NoStop}%
\bibitem [{\citenamefont {Rivest}\ \emph {et~al.}(1978)\citenamefont {Rivest},
  \citenamefont {Shamir},\ and\ \citenamefont {Adleman}}]{RSA78}%
  \BibitemOpen
  \bibfield  {author} {\bibinfo {author} {\bibfnamefont {R.}~\bibnamefont
  {Rivest}}, \bibinfo {author} {\bibfnamefont {A.}~\bibnamefont {Shamir}},\
  and\ \bibinfo {author} {\bibfnamefont {L.}~\bibnamefont {Adleman}},\
  }\bibfield  {title} {\bibinfo {title} {A method for obtaining digital
  signatures and public-key cryptosystems},\ }\href
  {https://doi.org/10.1145/359340.359342} {\bibfield  {journal} {\bibinfo
  {journal} {Commun. ACM}\ }\textbf {\bibinfo {volume} {21}},\ \bibinfo {pages}
  {120} (\bibinfo {year} {1978})}\BibitemShut {NoStop}%
\bibitem [{\citenamefont {Gueron}(2012)}]{Gue12}%
  \BibitemOpen
  \bibfield  {author} {\bibinfo {author} {\bibfnamefont {S.}~\bibnamefont
  {Gueron}},\ }\bibfield  {title} {\bibinfo {title} {Efficient software
  implementations of modular exponentiation},\ }\href
  {https://doi.org/10.1007/s13389-012-0031-5} {\bibfield  {journal} {\bibinfo
  {journal} {J. Cryptogr. Eng.}\ }\textbf {\bibinfo {volume} {2}},\ \bibinfo
  {pages} {31} (\bibinfo {year} {2012})}\BibitemShut {NoStop}%
\bibitem [{\citenamefont {Markov}\ and\ \citenamefont {Saeedi}(2012)}]{MS12}%
  \BibitemOpen
  \bibfield  {author} {\bibinfo {author} {\bibfnamefont {I.~L.}\ \bibnamefont
  {Markov}}\ and\ \bibinfo {author} {\bibfnamefont {M.}~\bibnamefont
  {Saeedi}},\ }\bibfield  {title} {\bibinfo {title} {Constant-optimized quantum
  circuits for modular multiplication and exponentiation},\ }\href
  {https://doi.org/10.26421/QIC12.5-6-1} {\bibfield  {journal} {\bibinfo
  {journal} {Quantum Inf. Comput.}\ }\textbf {\bibinfo {volume} {12}},\
  \bibinfo {pages} {0361} (\bibinfo {year} {2012})}\BibitemShut {NoStop}%
\bibitem [{\citenamefont {Sanders}(2023)}]{San23}%
  \BibitemOpen
  \bibfield  {author} {\bibinfo {author} {\bibfnamefont {Y.~R.}\ \bibnamefont
  {Sanders}},\ }\href@noop {} {} (\bibinfo {year} {2023}),\ \bibinfo {note}
  {private communication}\BibitemShut {NoStop}%
\bibitem [{\citenamefont {Boneh}\ and\ \citenamefont {Lipton}(1995)}]{BL95}%
  \BibitemOpen
  \bibfield  {author} {\bibinfo {author} {\bibfnamefont {D.}~\bibnamefont
  {Boneh}}\ and\ \bibinfo {author} {\bibfnamefont {R.~J.}\ \bibnamefont
  {Lipton}},\ }\bibfield  {title} {\bibinfo {title} {Quantum cryptanalysis of
  hidden linear functions},\ }in\ \href {https://doi.org/10.1007/3-540-44750-4}
  {\emph {\bibinfo {booktitle} {Advances in Cryptology --- CRYPTO '95}}},\
  \bibinfo {series} {Lecture Notes in Computer Science}, Vol.\ \bibinfo
  {volume} {963},\ \bibinfo {editor} {edited by\ \bibinfo {editor}
  {\bibfnamefont {D.}~\bibnamefont {Coppersmith}}}\ (\bibinfo  {publisher}
  {Springer-Verlag},\ \bibinfo {address} {Berlin},\ \bibinfo {year} {1995})\
  pp.\ \bibinfo {pages} {424--437}\BibitemShut {NoStop}%
\bibitem [{\citenamefont {Proos}\ and\ \citenamefont {Zalka}(2003)}]{PZ03}%
  \BibitemOpen
  \bibfield  {author} {\bibinfo {author} {\bibfnamefont {J.}~\bibnamefont
  {Proos}}\ and\ \bibinfo {author} {\bibfnamefont {C.}~\bibnamefont {Zalka}},\
  }\bibfield  {title} {\bibinfo {title} {Shor's discrete logarithm quantum
  algorithm for elliptic curves},\ }\href@noop {} {\bibfield  {journal}
  {\bibinfo  {journal} {Quantum Inf. Comput.}\ }\textbf {\bibinfo {volume}
  {3}},\ \bibinfo {pages} {317} (\bibinfo {year} {2003})}\BibitemShut {NoStop}%
\bibitem [{\citenamefont {Hallgren}(2002)}]{Hal02}%
  \BibitemOpen
  \bibfield  {author} {\bibinfo {author} {\bibfnamefont {S.}~\bibnamefont
  {Hallgren}},\ }\bibfield  {title} {\bibinfo {title} {Polynomial-time quantum
  algorithms for {P}ell's equation and the principal ideal problem},\ }in\
  \href@noop {} {\emph {\bibinfo {booktitle} {Proc.\ 34th ACM Symposium on
  Theory of Computing}}}\ (\bibinfo {year} {2002})\BibitemShut {NoStop}%
\bibitem [{\citenamefont {Buchmann}\ and\ \citenamefont
  {Williams}(1988)}]{BW88}%
  \BibitemOpen
  \bibfield  {author} {\bibinfo {author} {\bibfnamefont {J.}~\bibnamefont
  {Buchmann}}\ and\ \bibinfo {author} {\bibfnamefont {H.}~\bibnamefont
  {Williams}},\ }\bibfield  {title} {\bibinfo {title} {A key-exchange system
  based on imaginary quadratic fields},\ }\href@noop {} {\bibfield  {journal}
  {\bibinfo  {journal} {J. Cryptol.}\ }\textbf {\bibinfo {volume} {1}},\
  \bibinfo {pages} {107} (\bibinfo {year} {1988})}\BibitemShut {NoStop}%
\bibitem [{\citenamefont {Mosca}(2015)}]{Mos15}%
  \BibitemOpen
  \bibfield  {author} {\bibinfo {author} {\bibfnamefont {M.}~\bibnamefont
  {Mosca}},\ }\bibinfo {title} {Abelian hidden subgroup problem},\ in\ \href
  {https://doi.org/10.1007/978-3-642-27848-8_1-2} {\emph {\bibinfo {booktitle}
  {Encyclopedia of Algorithms}}},\ \bibinfo {editor} {edited by\ \bibinfo
  {editor} {\bibfnamefont {M.~Y.}\ \bibnamefont {Kao}}}\ (\bibinfo  {publisher}
  {Springer},\ \bibinfo {address} {Berlin},\ \bibinfo {year} {2015})\ pp.\
  \bibinfo {pages} {1--5}\BibitemShut {NoStop}%
\bibitem [{\citenamefont {Nielsen}\ and\ \citenamefont {Chuang}(2010)}]{NC10}%
  \BibitemOpen
  \bibfield  {author} {\bibinfo {author} {\bibfnamefont {M.~A.}\ \bibnamefont
  {Nielsen}}\ and\ \bibinfo {author} {\bibfnamefont {I.~L.}\ \bibnamefont
  {Chuang}},\ }\href@noop {} {\emph {\bibinfo {title} {Quantum {C}omputation
  and {Q}uantum {I}nformation: 10th Anniversary Edition}}}\ (\bibinfo
  {publisher} {Cambridge University Press},\ \bibinfo {address} {Cambridge},\
  \bibinfo {year} {2010})\BibitemShut {NoStop}%
\bibitem [{\citenamefont {de~Beaudrap}\ \emph {et~al.}(2002)\citenamefont
  {de~Beaudrap}, \citenamefont {Cleve},\ and\ \citenamefont
  {Watrous}}]{dBCW02}%
  \BibitemOpen
  \bibfield  {author} {\bibinfo {author} {\bibfnamefont {J.~N.}\ \bibnamefont
  {de~Beaudrap}}, \bibinfo {author} {\bibfnamefont {R.}~\bibnamefont {Cleve}},\
  and\ \bibinfo {author} {\bibfnamefont {J.}~\bibnamefont {Watrous}},\
  }\bibfield  {title} {\bibinfo {title} {Sharp quantum versus classical query
  complexity separations},\ }\href {https://doi.org/10.1007/s00453-002-0978-}
  {\bibfield  {journal} {\bibinfo  {journal} {Algorithmica}\ }\textbf {\bibinfo
  {volume} {34}},\ \bibinfo {pages} {449} (\bibinfo {year} {2002})}\BibitemShut
  {NoStop}%
\bibitem [{\citenamefont {Ivanyos}\ \emph {et~al.}(2001)\citenamefont
  {Ivanyos}, \citenamefont {Magniez},\ and\ \citenamefont {Santha}}]{IMS01}%
  \BibitemOpen
  \bibfield  {author} {\bibinfo {author} {\bibfnamefont {G.}~\bibnamefont
  {Ivanyos}}, \bibinfo {author} {\bibfnamefont {F.}~\bibnamefont {Magniez}},\
  and\ \bibinfo {author} {\bibfnamefont {M.}~\bibnamefont {Santha}},\
  }\bibfield  {title} {\bibinfo {title} {Efficient quantum algorithms for some
  instances of the non-abelian hidden subgroup problem},\ }in\ \href@noop {}
  {\emph {\bibinfo {booktitle} {SPAA '01: Proc.\ {t}hirteenth {A}nnual ACM
  {S}ymposium on {P}arallel {A}lgorithms and {A}rchitectures}}}\ (\bibinfo
  {publisher} {ACM},\ \bibinfo {address} {New York},\ \bibinfo {year} {2001})\
  pp.\ \bibinfo {pages} {263--270}\BibitemShut {NoStop}%
\bibitem [{\citenamefont {Ettinger}\ and\ \citenamefont
  {H{\o}yer}(1999)}]{EH99}%
  \BibitemOpen
  \bibfield  {author} {\bibinfo {author} {\bibfnamefont {M.}~\bibnamefont
  {Ettinger}}\ and\ \bibinfo {author} {\bibfnamefont {P.}~\bibnamefont
  {H{\o}yer}},\ }\href@noop {} {\bibinfo {title} {A quantum observable for the
  graph isomorphism problem}} (\bibinfo {year} {1999}),\ \Eprint
  {https://arxiv.org/abs/quant-ph/9901029} {arXiv:quant-ph/9901029 [quant-ph]}
  \BibitemShut {NoStop}%
\bibitem [{\citenamefont {Regev}(2002)}]{Reg02}%
  \BibitemOpen
  \bibfield  {author} {\bibinfo {author} {\bibfnamefont {O.}~\bibnamefont
  {Regev}},\ }\bibfield  {title} {\bibinfo {title} {Quantum computation and
  lattice problems},\ }in\ \href@noop {} {\emph {\bibinfo {booktitle} {Proc.
  3rd Symposium on Foundations of Computer Science}}}\ (\bibinfo {year}
  {2002})\ pp.\ \bibinfo {pages} {520--529}\BibitemShut {NoStop}%
\bibitem [{\citenamefont {Kuperberg}(2005)}]{Kup05}%
  \BibitemOpen
  \bibfield  {author} {\bibinfo {author} {\bibfnamefont {G.}~\bibnamefont
  {Kuperberg}},\ }\bibfield  {title} {\bibinfo {title} {A subexponential-time
  quantum algorithm for the dihedral hidden subgroup problem},\ }\href
  {https://doi.org/10.1137/S0097539703436345} {\bibfield  {journal} {\bibinfo
  {journal} {SIAM J. Comput.}\ }\textbf {\bibinfo {volume} {35}},\ \bibinfo
  {pages} {170} (\bibinfo {year} {2005})}\BibitemShut {NoStop}%
\bibitem [{\citenamefont {Regev}(2004)}]{Reg04}%
  \BibitemOpen
  \bibfield  {author} {\bibinfo {author} {\bibfnamefont {O.}~\bibnamefont
  {Regev}},\ }\href@noop {} {\bibinfo {title} {A subexponential time algorithm
  for the dihedral hidden subgroup problem with polynomial space}} (\bibinfo
  {year} {2004}),\ \Eprint {https://arxiv.org/abs/quant-ph/0406151}
  {arXiv:quant-ph/0406151 [quant-ph]} \BibitemShut {NoStop}%
\bibitem [{\citenamefont {Grover}(1997)}]{Gro97}%
  \BibitemOpen
  \bibfield  {author} {\bibinfo {author} {\bibfnamefont {L.}~\bibnamefont
  {Grover}},\ }\bibfield  {title} {\bibinfo {title} {Quantum mechanics helps in
  searching for a needle in a haystack},\ }\href@noop {} {\bibfield  {journal}
  {\bibinfo  {journal} {Phys. Rev. Lett.}\ }\textbf {\bibinfo {volume} {79}},\
  \bibinfo {pages} {325} (\bibinfo {year} {1997})}\BibitemShut {NoStop}%
\bibitem [{\citenamefont {Boyer}\ \emph {et~al.}(1998)\citenamefont {Boyer},
  \citenamefont {Brassard}, \citenamefont {H{\o}yer},\ and\ \citenamefont
  {Tapp}}]{BBHT98}%
  \BibitemOpen
  \bibfield  {author} {\bibinfo {author} {\bibfnamefont {M.}~\bibnamefont
  {Boyer}}, \bibinfo {author} {\bibfnamefont {G.}~\bibnamefont {Brassard}},
  \bibinfo {author} {\bibfnamefont {P.}~\bibnamefont {H{\o}yer}},\ and\
  \bibinfo {author} {\bibfnamefont {A.}~\bibnamefont {Tapp}},\ }\bibfield
  {title} {\bibinfo {title} {Tight bounds on quantum searching},\ }\href@noop
  {} {\bibfield  {journal} {\bibinfo  {journal} {Fortschritte der Physik}\
  }\textbf {\bibinfo {volume} {46}},\ \bibinfo {pages} {493} (\bibinfo {year}
  {1998})}\BibitemShut {NoStop}%
\bibitem [{\citenamefont {Bennett}\ \emph {et~al.}(1997)\citenamefont
  {Bennett}, \citenamefont {Bernstein}, \citenamefont {Brassard},\ and\
  \citenamefont {Vazirani}}]{BBBV97}%
  \BibitemOpen
  \bibfield  {author} {\bibinfo {author} {\bibfnamefont {C.~H.}\ \bibnamefont
  {Bennett}}, \bibinfo {author} {\bibfnamefont {E.}~\bibnamefont {Bernstein}},
  \bibinfo {author} {\bibfnamefont {G.}~\bibnamefont {Brassard}},\ and\
  \bibinfo {author} {\bibfnamefont {U.}~\bibnamefont {Vazirani}},\ }\bibfield
  {title} {\bibinfo {title} {Strengths and weaknesses of quantum computing},\
  }\href {https://doi.org/10.1137/S0097539796300933} {\bibfield  {journal}
  {\bibinfo  {journal} {SIAM J. Sci. Comput.}\ }\textbf {\bibinfo {volume}
  {26}},\ \bibinfo {pages} {1510} (\bibinfo {year} {1997})}\BibitemShut
  {NoStop}%
\bibitem [{\citenamefont {Zalka}(1999)}]{Zal99}%
  \BibitemOpen
  \bibfield  {author} {\bibinfo {author} {\bibfnamefont {C.}~\bibnamefont
  {Zalka}},\ }\bibfield  {title} {\bibinfo {title} {Grover's quantum searching
  algorithm is optimal},\ }\href {https://doi.org/10.1103/PhysRevA.60.2746}
  {\bibfield  {journal} {\bibinfo  {journal} {Phys. Rev. A}\ }\textbf {\bibinfo
  {volume} {60}},\ \bibinfo {pages} {2746} (\bibinfo {year}
  {1999})}\BibitemShut {NoStop}%
\bibitem [{\citenamefont {Brassard}\ and\ \citenamefont
  {H{\o}yer}(1997)}]{BH97}%
  \BibitemOpen
  \bibfield  {author} {\bibinfo {author} {\bibfnamefont {G.}~\bibnamefont
  {Brassard}}\ and\ \bibinfo {author} {\bibfnamefont {P.}~\bibnamefont
  {H{\o}yer}},\ }\bibfield  {title} {\bibinfo {title} {An exact quantum
  polynomial-time algorithm for simon's problem},\ }in\ \href@noop {} {\emph
  {\bibinfo {booktitle} {Proc. Fifth Israeli Symposium on Theory of Computing
  and Systems}}}\ (\bibinfo  {publisher} {IEEE Computer Society Press},\
  \bibinfo {year} {1997})\ pp.\ \bibinfo {pages} {12--23}\BibitemShut {NoStop}%
\bibitem [{\citenamefont {Grover}(1998)}]{Gro98}%
  \BibitemOpen
  \bibfield  {author} {\bibinfo {author} {\bibfnamefont {L.~K.}\ \bibnamefont
  {Grover}},\ }\bibfield  {title} {\bibinfo {title} {Quantum computers can
  search rapidly by using almost any transformation},\ }\href
  {https://doi.org/10.1103/PhysRevLett.80.4329} {\bibfield  {journal} {\bibinfo
   {journal} {Phys. Rev. Lett.}\ }\textbf {\bibinfo {volume} {80}},\ \bibinfo
  {pages} {4329} (\bibinfo {year} {1998})}\BibitemShut {NoStop}%
\bibitem [{\citenamefont {Brassard}\ \emph {et~al.}(2002)\citenamefont
  {Brassard}, \citenamefont {H{\o}yer}, \citenamefont {Mosca},\ and\
  \citenamefont {Tapp}}]{BHMT02}%
  \BibitemOpen
  \bibfield  {author} {\bibinfo {author} {\bibfnamefont {G.}~\bibnamefont
  {Brassard}}, \bibinfo {author} {\bibfnamefont {P.}~\bibnamefont {H{\o}yer}},
  \bibinfo {author} {\bibfnamefont {M.}~\bibnamefont {Mosca}},\ and\ \bibinfo
  {author} {\bibfnamefont {A.}~\bibnamefont {Tapp}},\ }\bibfield  {title}
  {\bibinfo {title} {Quantum amplitude amplification and estimation},\ }in\
  \href@noop {} {\emph {\bibinfo {booktitle} {Quantum Computation and Quantum
  Information: A Millennium Volume}}},\ \bibinfo {series} {AMS Contemporary
  Mathematics Series}, Vol.\ \bibinfo {volume} {305},\ \bibinfo {editor}
  {edited by\ \bibinfo {editor} {\bibfnamefont {S.~J.}\ \bibnamefont
  {Lomonaco~Jr.}}\ and\ \bibinfo {editor} {\bibfnamefont {H.~E.}\ \bibnamefont
  {Brandt}}}\ (\bibinfo  {publisher} {American Mathematical Society},\ \bibinfo
  {year} {2002})\BibitemShut {NoStop}%
\bibitem [{\citenamefont {Ambainis}(2004)}]{Amb04}%
  \BibitemOpen
  \bibfield  {author} {\bibinfo {author} {\bibfnamefont {A.}~\bibnamefont
  {Ambainis}},\ }\bibfield  {title} {\bibinfo {title} {Quantum search
  algorithms},\ }\href@noop {} {\bibfield  {journal} {\bibinfo  {journal}
  {SIGACT News}\ }\textbf {\bibinfo {volume} {35}},\ \bibinfo {pages} {22}
  (\bibinfo {year} {2004})}\BibitemShut {NoStop}%
\bibitem [{\citenamefont {Aharonov}\ and\ \citenamefont
  {Ta-Shma}(2003)}]{AT03}%
  \BibitemOpen
  \bibfield  {author} {\bibinfo {author} {\bibfnamefont {D.}~\bibnamefont
  {Aharonov}}\ and\ \bibinfo {author} {\bibfnamefont {A.}~\bibnamefont
  {Ta-Shma}},\ }\bibfield  {title} {\bibinfo {title} {Adiabatic quantum state
  generation and statistical zero knowledge},\ }in\ \href
  {https://doi.org/10.1145/780542.780546} {\emph {\bibinfo {booktitle} {Proc.\
  Thirty-Fifth Annual ACM Symposium on Theory of Computing}}},\ \bibinfo
  {series and number} {STOC '03}\ (\bibinfo  {publisher} {Association for
  Computing Machinery},\ \bibinfo {address} {New York},\ \bibinfo {year}
  {2003})\ pp.\ \bibinfo {pages} {20--29}\BibitemShut {NoStop}%
\bibitem [{\citenamefont {Feynman}(1982)}]{Fey82}%
  \BibitemOpen
  \bibfield  {author} {\bibinfo {author} {\bibfnamefont {R.~P.}\ \bibnamefont
  {Feynman}},\ }\bibfield  {title} {\bibinfo {title} {Simulating physics with
  computers},\ }\href@noop {} {\bibfield  {journal} {\bibinfo  {journal} {Int.
  J. Theor. Phys.}\ }\textbf {\bibinfo {volume} {21}},\ \bibinfo {pages} {467}
  (\bibinfo {year} {1982})}\BibitemShut {NoStop}%
\bibitem [{\citenamefont {Kassal}\ \emph {et~al.}(2008)\citenamefont {Kassal},
  \citenamefont {Jordan}, \citenamefont {Love}, \citenamefont {Mohseni},\ and\
  \citenamefont {Aspuru-Guzik}}]{KJL+08}%
  \BibitemOpen
  \bibfield  {author} {\bibinfo {author} {\bibfnamefont {I.}~\bibnamefont
  {Kassal}}, \bibinfo {author} {\bibfnamefont {S.~P.}\ \bibnamefont {Jordan}},
  \bibinfo {author} {\bibfnamefont {P.~J.}\ \bibnamefont {Love}}, \bibinfo
  {author} {\bibfnamefont {M.}~\bibnamefont {Mohseni}},\ and\ \bibinfo {author}
  {\bibfnamefont {A.}~\bibnamefont {Aspuru-Guzik}},\ }\bibfield  {title}
  {\bibinfo {title} {Polynomial-time quantum algorithm for the simulation of
  chemical dynamics},\ }\href {https://doi.org/10.1073/pnas.0808245105}
  {\bibfield  {journal} {\bibinfo  {journal} {Proc. Natl. Acad. Sci. U.S.A.}\
  }\textbf {\bibinfo {volume} {105}},\ \bibinfo {pages} {18681} (\bibinfo
  {year} {2008})}\BibitemShut {NoStop}%
\bibitem [{\citenamefont {Childs}(2004)}]{Chi04}%
  \BibitemOpen
  \bibfield  {author} {\bibinfo {author} {\bibfnamefont {A.~M.}\ \bibnamefont
  {Childs}},\ }\emph {\bibinfo {title} {Quantum information processing in
  continuous time}},\ \href@noop {} {Ph.D. thesis},\ \bibinfo  {school} {MIT}
  (\bibinfo {year} {2004})\BibitemShut {NoStop}%
\bibitem [{\citenamefont {Berry}\ \emph
  {et~al.}(2007{\natexlab{a}})\citenamefont {Berry}, \citenamefont {Ahokas},
  \citenamefont {Cleve},\ and\ \citenamefont {Sanders}}]{BACS07}%
  \BibitemOpen
  \bibfield  {author} {\bibinfo {author} {\bibfnamefont {D.~W.}\ \bibnamefont
  {Berry}}, \bibinfo {author} {\bibfnamefont {G.}~\bibnamefont {Ahokas}},
  \bibinfo {author} {\bibfnamefont {R.}~\bibnamefont {Cleve}},\ and\ \bibinfo
  {author} {\bibfnamefont {B.~C.}\ \bibnamefont {Sanders}},\ }\bibfield
  {title} {\bibinfo {title} {Efficient quantum algorithms for simulating sparse
  {H}amiltonians},\ }\href@noop {} {\bibfield  {journal} {\bibinfo  {journal}
  {Commun. Math. Phys}\ }\textbf {\bibinfo {volume} {270}},\ \bibinfo {pages}
  {359} (\bibinfo {year} {2007}{\natexlab{a}})}\BibitemShut {NoStop}%
\bibitem [{\citenamefont {Berry}\ \emph
  {et~al.}(2007{\natexlab{b}})\citenamefont {Berry}, \citenamefont {Ahokas},
  \citenamefont {Cleve},\ and\ \citenamefont {Sanders}}]{BACS07ch}%
  \BibitemOpen
  \bibfield  {author} {\bibinfo {author} {\bibfnamefont {D.~W.}\ \bibnamefont
  {Berry}}, \bibinfo {author} {\bibfnamefont {G.}~\bibnamefont {Ahokas}},
  \bibinfo {author} {\bibfnamefont {R.}~\bibnamefont {Cleve}},\ and\ \bibinfo
  {author} {\bibfnamefont {B.~C.}\ \bibnamefont {Sanders}},\ }\bibfield
  {title} {\bibinfo {title} {Quantum algorithms for {H}amiltonian simulation},\
  }in\ \href@noop {} {\emph {\bibinfo {booktitle} {Mathematics of Quantum
  Computation and Quantum Technology}}},\ \bibinfo {editor} {edited by\
  \bibinfo {editor} {\bibfnamefont {G.}~\bibnamefont {Chen}}, \bibinfo {editor}
  {\bibfnamefont {L.}~\bibnamefont {Kauffman}},\ and\ \bibinfo {editor}
  {\bibfnamefont {S.~J.}\ \bibnamefont {Lomonaco}}}\ (\bibinfo  {publisher}
  {Taylor \&\ Francis},\ \bibinfo {address} {Oxford UK},\ \bibinfo {year}
  {2007})\ pp.\ \bibinfo {pages} {89--110}\BibitemShut {NoStop}%
\bibitem [{\citenamefont {Kempe}\ \emph {et~al.}(2004)\citenamefont {Kempe},
  \citenamefont {Kitaev},\ and\ \citenamefont {Regev}}]{KKR04}%
  \BibitemOpen
  \bibfield  {author} {\bibinfo {author} {\bibfnamefont {J.}~\bibnamefont
  {Kempe}}, \bibinfo {author} {\bibfnamefont {A.}~\bibnamefont {Kitaev}},\ and\
  \bibinfo {author} {\bibfnamefont {O.}~\bibnamefont {Regev}},\ }\bibfield
  {title} {\bibinfo {title} {The complexity of the local {H}amiltonian
  problem},\ }in\ \href@noop {} {\emph {\bibinfo {booktitle} {Proc. 24th
  FSTTCS}}},\ \bibinfo {series} {LNCS 3328}, Vol.~\bibinfo {volume} {35}\
  (\bibinfo  {publisher} {Springer-Verlag},\ \bibinfo {address} {Berlin},\
  \bibinfo {year} {2004})\ pp.\ \bibinfo {pages} {372--383}\BibitemShut
  {NoStop}%
\bibitem [{\citenamefont {Borchers}\ and\ \citenamefont {Furman}(1998)}]{BF98}%
  \BibitemOpen
  \bibfield  {author} {\bibinfo {author} {\bibfnamefont {B.}~\bibnamefont
  {Borchers}}\ and\ \bibinfo {author} {\bibfnamefont {J.}~\bibnamefont
  {Furman}},\ }\bibfield  {title} {\bibinfo {title} {A two-phase exact
  algorithm for {MAX}-{SAT} and {W}eighted {MAX}-{SAT} problems},\ }\href
  {https://doi.org/10.1023/A:1009725216438} {\bibfield  {journal} {\bibinfo
  {journal} {J. Comb. Optim.}\ }\textbf {\bibinfo {volume} {2}},\ \bibinfo
  {pages} {299} (\bibinfo {year} {1998})}\BibitemShut {NoStop}%
\bibitem [{\citenamefont {Harrow}\ \emph {et~al.}(2009)\citenamefont {Harrow},
  \citenamefont {Hassidim},\ and\ \citenamefont {Lloyd}}]{HHL09}%
  \BibitemOpen
  \bibfield  {author} {\bibinfo {author} {\bibfnamefont {A.~W.}\ \bibnamefont
  {Harrow}}, \bibinfo {author} {\bibfnamefont {A.}~\bibnamefont {Hassidim}},\
  and\ \bibinfo {author} {\bibfnamefont {S.}~\bibnamefont {Lloyd}},\ }\bibfield
   {title} {\bibinfo {title} {Quantum algorithm for linear systems of
  equations},\ }\href {https://doi.org/10.1103/PhysRevLett.103.150502}
  {\bibfield  {journal} {\bibinfo  {journal} {Phys. Rev. Lett.}\ }\textbf
  {\bibinfo {volume} {103}},\ \bibinfo {pages} {150502} (\bibinfo {year}
  {2009})}\BibitemShut {NoStop}%
\bibitem [{\citenamefont {Alase}\ \emph {et~al.}(2022)\citenamefont {Alase},
  \citenamefont {Nerem}, \citenamefont {Bagherimehrab}, \citenamefont
  {H\o{}yer},\ and\ \citenamefont {Sanders}}]{ANB+22}%
  \BibitemOpen
  \bibfield  {author} {\bibinfo {author} {\bibfnamefont {A.}~\bibnamefont
  {Alase}}, \bibinfo {author} {\bibfnamefont {R.~R.}\ \bibnamefont {Nerem}},
  \bibinfo {author} {\bibfnamefont {M.}~\bibnamefont {Bagherimehrab}}, \bibinfo
  {author} {\bibfnamefont {P.}~\bibnamefont {H\o{}yer}},\ and\ \bibinfo
  {author} {\bibfnamefont {B.~C.}\ \bibnamefont {Sanders}},\ }\bibfield
  {title} {\bibinfo {title} {Tight bound for estimating expectation values from
  a system of linear equations},\ }\href
  {https://doi.org/10.1103/PhysRevResearch.4.023237} {\bibfield  {journal}
  {\bibinfo  {journal} {Phys. Rev. Res.}\ }\textbf {\bibinfo {volume} {4}},\
  \bibinfo {pages} {023237} (\bibinfo {year} {2022})}\BibitemShut {NoStop}%
\bibitem [{\citenamefont {Belsley}\ \emph {et~al.}(2005)\citenamefont
  {Belsley}, \citenamefont {Kuh},\ and\ \citenamefont {Welsch}}]{BKW05}%
  \BibitemOpen
  \bibfield  {author} {\bibinfo {author} {\bibfnamefont {D.~A.}\ \bibnamefont
  {Belsley}}, \bibinfo {author} {\bibfnamefont {E.}~\bibnamefont {Kuh}},\ and\
  \bibinfo {author} {\bibfnamefont {R.~E.}\ \bibnamefont {Welsch}},\
  }\href@noop {} {\emph {\bibinfo {title} {Regression Diagnostics: Identifying
  Influential Data and Sources of Collinearity}}}\ (\bibinfo  {publisher}
  {Wiley},\ \bibinfo {address} {New York},\ \bibinfo {year} {2005})\
  Chap.~\bibinfo {chapter} {3}, pp.\ \bibinfo {pages} {100--105}\BibitemShut
  {NoStop}%
\bibitem [{\citenamefont {Kerenidis}\ and\ \citenamefont
  {Prakash}(2017)}]{KP17}%
  \BibitemOpen
  \bibfield  {author} {\bibinfo {author} {\bibfnamefont {I.}~\bibnamefont
  {Kerenidis}}\ and\ \bibinfo {author} {\bibfnamefont {A.}~\bibnamefont
  {Prakash}},\ }\bibfield  {title} {\bibinfo {title} {Quantum recommendation
  system},\ }in\ \href@noop {} {\emph {\bibinfo {booktitle} {Innovations in
  Theoretical Computer Science (ITCS 2017)}}},\ \bibinfo {series} {LIPIcs},
  Vol.~\bibinfo {volume} {67}\ (\bibinfo {year} {2017})\ pp.\ \bibinfo {pages}
  {1868--8969}\BibitemShut {NoStop}%
\bibitem [{\citenamefont {Wossnig}\ \emph {et~al.}(2018)\citenamefont
  {Wossnig}, \citenamefont {Zhao},\ and\ \citenamefont {Prakash}}]{WZP18}%
  \BibitemOpen
  \bibfield  {author} {\bibinfo {author} {\bibfnamefont {L.}~\bibnamefont
  {Wossnig}}, \bibinfo {author} {\bibfnamefont {Z.}~\bibnamefont {Zhao}},\ and\
  \bibinfo {author} {\bibfnamefont {A.}~\bibnamefont {Prakash}},\ }\bibfield
  {title} {\bibinfo {title} {Quantum linear system algorithm for dense
  matrices},\ }\href {https://doi.org/10.1103/PhysRevLett.120.050502}
  {\bibfield  {journal} {\bibinfo  {journal} {Phys. Rev. Lett.}\ }\textbf
  {\bibinfo {volume} {120}},\ \bibinfo {pages} {050502} (\bibinfo {year}
  {2018})}\BibitemShut {NoStop}%
\bibitem [{\citenamefont {Lloyd}\ \emph {et~al.}(2014)\citenamefont {Lloyd},
  \citenamefont {Mohseni},\ and\ \citenamefont {Rebentrost}}]{LMR14}%
  \BibitemOpen
  \bibfield  {author} {\bibinfo {author} {\bibfnamefont {S.}~\bibnamefont
  {Lloyd}}, \bibinfo {author} {\bibfnamefont {M.}~\bibnamefont {Mohseni}},\
  and\ \bibinfo {author} {\bibfnamefont {P.}~\bibnamefont {Rebentrost}},\
  }\bibfield  {title} {\bibinfo {title} {Quantum principal component
  analysis},\ }\href {https://doi.org/10.1038/nphys3029} {\bibfield  {journal}
  {\bibinfo  {journal} {Nat. Phys.}\ }\textbf {\bibinfo {volume} {10}},\
  \bibinfo {pages} {631} (\bibinfo {year} {2014})}\BibitemShut {NoStop}%
\bibitem [{\citenamefont {Tang}(2019)}]{Tan19}%
  \BibitemOpen
  \bibfield  {author} {\bibinfo {author} {\bibfnamefont {E.}~\bibnamefont
  {Tang}},\ }\bibfield  {title} {\bibinfo {title} {A quantum-inspired classical
  algorithm for recommendation systems},\ }in\ \href@noop {} {\emph {\bibinfo
  {booktitle} {Proc. 51st Annual ACM SIGACT Symposium on Theory of
  Computing}}}\ (\bibinfo {year} {2019})\ pp.\ \bibinfo {pages}
  {217--228}\BibitemShut {NoStop}%
\bibitem [{\citenamefont {Tang}(2021)}]{Tan21}%
  \BibitemOpen
  \bibfield  {author} {\bibinfo {author} {\bibfnamefont {E.}~\bibnamefont
  {Tang}},\ }\bibfield  {title} {\bibinfo {title} {Quantum principal component
  analysis only achieves an exponential speedup because of its state
  preparation assumptions},\ }\href
  {https://doi.org/10.1103/PhysRevLett.127.060503} {\bibfield  {journal}
  {\bibinfo  {journal} {Phys. Rev. Lett.}\ }\textbf {\bibinfo {volume} {127}},\
  \bibinfo {pages} {060503} (\bibinfo {year} {2021})}\BibitemShut {NoStop}%
\bibitem [{\citenamefont {Gily\'{e}n}\ \emph {et~al.}(2019)\citenamefont
  {Gily\'{e}n}, \citenamefont {Su}, \citenamefont {Low},\ and\ \citenamefont
  {Wiebe}}]{GSLW19}%
  \BibitemOpen
  \bibfield  {author} {\bibinfo {author} {\bibfnamefont {A.}~\bibnamefont
  {Gily\'{e}n}}, \bibinfo {author} {\bibfnamefont {Y.}~\bibnamefont {Su}},
  \bibinfo {author} {\bibfnamefont {G.~H.}\ \bibnamefont {Low}},\ and\ \bibinfo
  {author} {\bibfnamefont {N.}~\bibnamefont {Wiebe}},\ }\bibfield  {title}
  {\bibinfo {title} {Quantum singular value transformation and beyond:
  exponential improvements for quantum matrix arithmetics},\ }in\ \href
  {https://doi.org/10.1145/3313276.3316366} {\emph {\bibinfo {booktitle} {Proc.
  51st Annual ACM SIGACT Symposium on Theory of Computing (STOC 2019), pp.
  193-204}}}\ (\bibinfo {year} {2019})\ pp.\ \bibinfo {pages}
  {193--204}\BibitemShut {NoStop}%
\bibitem [{\citenamefont {Chakraborty}\ \emph {et~al.}(2019)\citenamefont
  {Chakraborty}, \citenamefont {Gily\'{e}n},\ and\ \citenamefont
  {Jeffery}}]{CGJ19}%
  \BibitemOpen
  \bibfield  {author} {\bibinfo {author} {\bibfnamefont {S.}~\bibnamefont
  {Chakraborty}}, \bibinfo {author} {\bibfnamefont {A.}~\bibnamefont
  {Gily\'{e}n}},\ and\ \bibinfo {author} {\bibfnamefont {S.}~\bibnamefont
  {Jeffery}},\ }\bibfield  {title} {\bibinfo {title} {{The Power of
  Block-Encoded Matrix Powers: Improved Regression Techniques via Faster
  Hamiltonian Simulation}},\ }in\ \href
  {https://doi.org/10.4230/LIPIcs.ICALP.2019.33} {\emph {\bibinfo {booktitle}
  {46th International Colloquium on Automata, Languages, and Programming (ICALP
  2019)}}},\ \bibinfo {series} {Leibniz International Proceedings in
  Informatics (LIPIcs)}, Vol.\ \bibinfo {volume} {132},\ \bibinfo {editor}
  {edited by\ \bibinfo {editor} {\bibfnamefont {C.}~\bibnamefont {Baier}},
  \bibinfo {editor} {\bibfnamefont {I.}~\bibnamefont {Chatzigiannakis}},
  \bibinfo {editor} {\bibfnamefont {P.}~\bibnamefont {Flocchini}},\ and\
  \bibinfo {editor} {\bibfnamefont {S.}~\bibnamefont {Leonardi}}}\ (\bibinfo
  {publisher} {Schloss Dagstuhl -- Leibniz-Zentrum f{\"u}r Informatik},\
  \bibinfo {address} {Dagstuhl},\ \bibinfo {year} {2019})\ pp.\ \bibinfo
  {pages} {33:1--33:14}\BibitemShut {NoStop}%
\bibitem [{\citenamefont {Jethwani}\ \emph {et~al.}(2020)\citenamefont
  {Jethwani}, \citenamefont {Le~Gall},\ and\ \citenamefont {Singh}}]{JLS20}%
  \BibitemOpen
  \bibfield  {author} {\bibinfo {author} {\bibfnamefont {D.}~\bibnamefont
  {Jethwani}}, \bibinfo {author} {\bibfnamefont {F.}~\bibnamefont {Le~Gall}},\
  and\ \bibinfo {author} {\bibfnamefont {S.~K.}\ \bibnamefont {Singh}},\
  }\bibfield  {title} {\bibinfo {title} {Quantum-inspired classical algorithms
  for singular value transformation},\ }in\ \href@noop {} {\emph {\bibinfo
  {booktitle} {Proc.\ 45th International Symposium on Mathematical Foundations
  of Computer Science (MFCS 2020), 53:1-53:14}}}\ (\bibinfo {year}
  {2020})\BibitemShut {NoStop}%
\bibitem [{\citenamefont {Farhi}\ \emph {et~al.}(2014)\citenamefont {Farhi},
  \citenamefont {Goldstone},\ and\ \citenamefont {Gutmann}}]{FGG14}%
  \BibitemOpen
  \bibfield  {author} {\bibinfo {author} {\bibfnamefont {E.}~\bibnamefont
  {Farhi}}, \bibinfo {author} {\bibfnamefont {J.}~\bibnamefont {Goldstone}},\
  and\ \bibinfo {author} {\bibfnamefont {S.}~\bibnamefont {Gutmann}},\
  }\href@noop {} {\bibinfo {title} {A quantum approximate optimization
  algorithm}} (\bibinfo {year} {2014}),\ \Eprint
  {https://arxiv.org/abs/1411.4028} {arXiv:1411.4028 [quant-ph]} \BibitemShut
  {NoStop}%
\bibitem [{\citenamefont {Farhi}\ \emph {et~al.}(2022)\citenamefont {Farhi},
  \citenamefont {Goldstone}, \citenamefont {Gutmann},\ and\ \citenamefont
  {Zhou}}]{FGGZ22}%
  \BibitemOpen
  \bibfield  {author} {\bibinfo {author} {\bibfnamefont {E.}~\bibnamefont
  {Farhi}}, \bibinfo {author} {\bibfnamefont {J.}~\bibnamefont {Goldstone}},
  \bibinfo {author} {\bibfnamefont {S.}~\bibnamefont {Gutmann}},\ and\ \bibinfo
  {author} {\bibfnamefont {L.}~\bibnamefont {Zhou}},\ }\bibfield  {title}
  {\bibinfo {title} {The {Q}uantum {A}pproximate {O}ptimization {A}lgorithm and
  the {S}herrington-{K}irkpatrick {M}odel at {I}nfinite {S}ize},\ }\href
  {https://doi.org/10.22331/q-2022-07-07-759} {\bibfield  {journal} {\bibinfo
  {journal} {{Quantum}}\ }\textbf {\bibinfo {volume} {6}},\ \bibinfo {pages}
  {759} (\bibinfo {year} {2022})}\BibitemShut {NoStop}%
\bibitem [{\citenamefont {Farhi}\ \emph {et~al.}(2015)\citenamefont {Farhi},
  \citenamefont {Goldstone},\ and\ \citenamefont {Gutmann}}]{FGG15}%
  \BibitemOpen
  \bibfield  {author} {\bibinfo {author} {\bibfnamefont {E.}~\bibnamefont
  {Farhi}}, \bibinfo {author} {\bibfnamefont {J.}~\bibnamefont {Goldstone}},\
  and\ \bibinfo {author} {\bibfnamefont {S.}~\bibnamefont {Gutmann}},\
  }\href@noop {} {\bibinfo {title} {A quantum approximate optimization
  algorithm applied to a bounded occurrence constraint problem}} (\bibinfo
  {year} {2015}),\ \Eprint {https://arxiv.org/abs/1412.6062} {arXiv:1412.6062
  [quant-ph]} \BibitemShut {NoStop}%
\bibitem [{\citenamefont {Barak}\ \emph {et~al.}(2015)\citenamefont {Barak},
  \citenamefont {Moitra}, \citenamefont {O’Donnell}, \citenamefont
  {Raghavendra}, \citenamefont {Regev}, \citenamefont {Steurer}, \citenamefont
  {Trevisan}, \citenamefont {Vijayaraghavan}, \citenamefont {Witmer},\ and\
  \citenamefont {Wright}}]{BMO+15}%
  \BibitemOpen
  \bibfield  {author} {\bibinfo {author} {\bibfnamefont {B.}~\bibnamefont
  {Barak}}, \bibinfo {author} {\bibfnamefont {A.}~\bibnamefont {Moitra}},
  \bibinfo {author} {\bibfnamefont {R.}~\bibnamefont {O’Donnell}}, \bibinfo
  {author} {\bibfnamefont {P.}~\bibnamefont {Raghavendra}}, \bibinfo {author}
  {\bibfnamefont {O.}~\bibnamefont {Regev}}, \bibinfo {author} {\bibfnamefont
  {D.}~\bibnamefont {Steurer}}, \bibinfo {author} {\bibfnamefont
  {L.}~\bibnamefont {Trevisan}}, \bibinfo {author} {\bibfnamefont
  {A.}~\bibnamefont {Vijayaraghavan}}, \bibinfo {author} {\bibfnamefont
  {D.}~\bibnamefont {Witmer}},\ and\ \bibinfo {author} {\bibfnamefont
  {J.}~\bibnamefont {Wright}},\ }\bibfield  {title} {\bibinfo {title} {Beating
  the random assignment on constraint satisfaction problems of bounded
  degree},\ }in\ \href {https://doi.org/10.4230/LIPIcs.APPROX-RANDOM.2015.110}
  {\emph {\bibinfo {booktitle} {Approximation, Randomization, and Combinatorial
  Optimization. Algorithms and Techniques (APPROX/RANDOM 2015)}}},\ \bibinfo
  {series} {Leibniz International Proceedings in Informatics (LIPIcs)},
  Vol.~\bibinfo {volume} {40},\ \bibinfo {editor} {edited by\ \bibinfo {editor}
  {\bibfnamefont {N.}~\bibnamefont {Garg}}, \bibinfo {editor} {\bibfnamefont
  {K.}~\bibnamefont {Jansen}}, \bibinfo {editor} {\bibfnamefont
  {A.}~\bibnamefont {Rao}},\ and\ \bibinfo {editor} {\bibfnamefont {J.~D.~P.}\
  \bibnamefont {Rolim}}}\ (\bibinfo  {publisher} {Schloss Dagstuhl --
  Leibniz-Zentrum f{\"u}r Informatik},\ \bibinfo {address} {Dagstuhl},\
  \bibinfo {year} {2015})\ pp.\ \bibinfo {pages} {110--123}\BibitemShut
  {NoStop}%
\bibitem [{\citenamefont {Raussendorf}\ and\ \citenamefont
  {Briegel}(2001)}]{RB01}%
  \BibitemOpen
  \bibfield  {author} {\bibinfo {author} {\bibfnamefont {R.}~\bibnamefont
  {Raussendorf}}\ and\ \bibinfo {author} {\bibfnamefont {H.~J.}\ \bibnamefont
  {Briegel}},\ }\bibfield  {title} {\bibinfo {title} {A one-way quantum
  computer},\ }\href {https://doi.org/10.1103/PhysRevLett.86.5188} {\bibfield
  {journal} {\bibinfo  {journal} {Phys. Rev. Lett.}\ }\textbf {\bibinfo
  {volume} {86}},\ \bibinfo {pages} {5188} (\bibinfo {year}
  {2001})}\BibitemShut {NoStop}%
\bibitem [{\citenamefont {Childs}\ \emph {et~al.}(2013)\citenamefont {Childs},
  \citenamefont {Gosset},\ and\ \citenamefont {Webb}}]{CGW13}%
  \BibitemOpen
  \bibfield  {author} {\bibinfo {author} {\bibfnamefont {A.~M.}\ \bibnamefont
  {Childs}}, \bibinfo {author} {\bibfnamefont {D.}~\bibnamefont {Gosset}},\
  and\ \bibinfo {author} {\bibfnamefont {Z.}~\bibnamefont {Webb}},\ }\bibfield
  {title} {\bibinfo {title} {Universal computation by multiparticle quantum
  walk},\ }\href {https://doi.org/10.1126/science.1229957} {\bibfield
  {journal} {\bibinfo  {journal} {Science}\ }\textbf {\bibinfo {volume}
  {339}},\ \bibinfo {pages} {791} (\bibinfo {year} {2013})}\BibitemShut
  {NoStop}%
\bibitem [{\citenamefont {Kitaev}(2003)}]{Kit03}%
  \BibitemOpen
  \bibfield  {author} {\bibinfo {author} {\bibfnamefont {A.}~\bibnamefont
  {Kitaev}},\ }\bibfield  {title} {\bibinfo {title} {Fault-tolerant quantum
  computation by anyons},\ }\href
  {https://doi.org/10.1016/S0003-4916(02)00018-0} {\bibfield  {journal}
  {\bibinfo  {journal} {Ann. Phys. (N.Y.)}\ }\textbf {\bibinfo {volume}
  {303}},\ \bibinfo {pages} {2} (\bibinfo {year} {2003})}\BibitemShut {NoStop}%
\bibitem [{\citenamefont {Farhi}\ \emph {et~al.}(2000)\citenamefont {Farhi},
  \citenamefont {Goldstone}, \citenamefont {Gutmann},\ and\ \citenamefont
  {Sipser}}]{FGGS00}%
  \BibitemOpen
  \bibfield  {author} {\bibinfo {author} {\bibfnamefont {E.}~\bibnamefont
  {Farhi}}, \bibinfo {author} {\bibfnamefont {J.}~\bibnamefont {Goldstone}},
  \bibinfo {author} {\bibfnamefont {S.}~\bibnamefont {Gutmann}},\ and\ \bibinfo
  {author} {\bibfnamefont {M.}~\bibnamefont {Sipser}},\ }\href@noop {}
  {\bibinfo {title} {Quantum computation by adiabatic evolution}} (\bibinfo
  {year} {2000}),\ \Eprint {https://arxiv.org/abs/quant-ph/0001106}
  {arXiv:quant-ph/0001106 [quant-ph]} \BibitemShut {NoStop}%
\bibitem [{\citenamefont {Farhi}\ \emph {et~al.}(2001)\citenamefont {Farhi},
  \citenamefont {Goldstone}, \citenamefont {Gutmann}, \citenamefont {Lapan},
  \citenamefont {Lundgren},\ and\ \citenamefont {Preda}}]{FGG+01}%
  \BibitemOpen
  \bibfield  {author} {\bibinfo {author} {\bibfnamefont {E.}~\bibnamefont
  {Farhi}}, \bibinfo {author} {\bibfnamefont {J.}~\bibnamefont {Goldstone}},
  \bibinfo {author} {\bibfnamefont {S.}~\bibnamefont {Gutmann}}, \bibinfo
  {author} {\bibfnamefont {J.}~\bibnamefont {Lapan}}, \bibinfo {author}
  {\bibfnamefont {A.}~\bibnamefont {Lundgren}},\ and\ \bibinfo {author}
  {\bibfnamefont {D.}~\bibnamefont {Preda}},\ }\bibfield  {title} {\bibinfo
  {title} {A quantum adiabatic evolution algorithm applied to random instances
  of an {NP}-complete problem},\ }\href
  {https://doi.org/10.1126/science.1057726} {\bibfield  {journal} {\bibinfo
  {journal} {Science}\ }\textbf {\bibinfo {volume} {292}},\ \bibinfo {pages}
  {472} (\bibinfo {year} {2001})}\BibitemShut {NoStop}%
\bibitem [{\citenamefont {Marzlin}\ and\ \citenamefont {Sanders}(2004)}]{MS04}%
  \BibitemOpen
  \bibfield  {author} {\bibinfo {author} {\bibfnamefont {K.-P.}\ \bibnamefont
  {Marzlin}}\ and\ \bibinfo {author} {\bibfnamefont {B.~C.}\ \bibnamefont
  {Sanders}},\ }\bibfield  {title} {\bibinfo {title} {Inconsistency in the
  application of the adiabatic theorem},\ }\href
  {https://doi.org/10.1103/PhysRevLett.93.160408} {\bibfield  {journal}
  {\bibinfo  {journal} {Phys. Rev. Lett.}\ }\textbf {\bibinfo {volume} {93}},\
  \bibinfo {pages} {160408} (\bibinfo {year} {2004})}\BibitemShut {NoStop}%
\bibitem [{\citenamefont {Jansen}\ \emph {et~al.}(2007)\citenamefont {Jansen},
  \citenamefont {Ruskai},\ and\ \citenamefont {Seiler}}]{JRS07}%
  \BibitemOpen
  \bibfield  {author} {\bibinfo {author} {\bibfnamefont {S.}~\bibnamefont
  {Jansen}}, \bibinfo {author} {\bibfnamefont {M.-B.}\ \bibnamefont {Ruskai}},\
  and\ \bibinfo {author} {\bibfnamefont {R.}~\bibnamefont {Seiler}},\
  }\bibfield  {title} {\bibinfo {title} {{Bounds for the adiabatic
  approximation with applications to quantum computation}},\ }\href
  {https://doi.org/10.1063/1.2798382} {\bibfield  {journal} {\bibinfo
  {journal} {J. Math. Phys.}\ }\textbf {\bibinfo {volume} {48}},\ \bibinfo
  {pages} {102111} (\bibinfo {year} {2007})}\BibitemShut {NoStop}%
\bibitem [{\citenamefont {Elgart}\ and\ \citenamefont {Hagedorn}(2012)}]{EH12}%
  \BibitemOpen
  \bibfield  {author} {\bibinfo {author} {\bibfnamefont {A.}~\bibnamefont
  {Elgart}}\ and\ \bibinfo {author} {\bibfnamefont {G.~A.}\ \bibnamefont
  {Hagedorn}},\ }\bibfield  {title} {\bibinfo {title} {{A note on the switching
  adiabatic theorem}},\ }\href {https://doi.org/10.1063/1.4748968} {\bibfield
  {journal} {\bibinfo  {journal} {J. Math. Phys.}\ }\textbf {\bibinfo {volume}
  {53}},\ \bibinfo {pages} {102202} (\bibinfo {year} {2012})}\BibitemShut
  {NoStop}%
\bibitem [{\citenamefont {Bravyi}\ \emph {et~al.}(2008)\citenamefont {Bravyi},
  \citenamefont {DiVincenzo}, \citenamefont {Oliveira},\ and\ \citenamefont
  {Terhal}}]{BDOT08}%
  \BibitemOpen
  \bibfield  {author} {\bibinfo {author} {\bibfnamefont {S.}~\bibnamefont
  {Bravyi}}, \bibinfo {author} {\bibfnamefont {D.~P.}\ \bibnamefont
  {DiVincenzo}}, \bibinfo {author} {\bibfnamefont {R.~I.}\ \bibnamefont
  {Oliveira}},\ and\ \bibinfo {author} {\bibfnamefont {B.~M.}\ \bibnamefont
  {Terhal}},\ }\bibfield  {title} {\bibinfo {title} {The complexity of
  stoquastic local {H}amiltonian problems},\ }\href@noop {} {\bibfield
  {journal} {\bibinfo  {journal} {Quant. Inf. Comp.}\ }\textbf {\bibinfo
  {volume} {8}},\ \bibinfo {pages} {0361} (\bibinfo {year} {2008})}\BibitemShut
  {NoStop}%
\bibitem [{\citenamefont {Aharonov}\ \emph {et~al.}(2007)\citenamefont
  {Aharonov}, \citenamefont {van Dam}, \citenamefont {Kempe}, \citenamefont
  {Landau}, \citenamefont {Lloyd},\ and\ \citenamefont {Regev}}]{AvDK+07}%
  \BibitemOpen
  \bibfield  {author} {\bibinfo {author} {\bibfnamefont {D.}~\bibnamefont
  {Aharonov}}, \bibinfo {author} {\bibfnamefont {W.}~\bibnamefont {van Dam}},
  \bibinfo {author} {\bibfnamefont {J.}~\bibnamefont {Kempe}}, \bibinfo
  {author} {\bibfnamefont {Z.}~\bibnamefont {Landau}}, \bibinfo {author}
  {\bibfnamefont {S.}~\bibnamefont {Lloyd}},\ and\ \bibinfo {author}
  {\bibfnamefont {O.}~\bibnamefont {Regev}},\ }\bibfield  {title} {\bibinfo
  {title} {Adiabatic quantum computation is equivalent to standard quantum
  computation},\ }\href@noop {} {\bibfield  {journal} {\bibinfo  {journal}
  {SIAM J. Comput.}\ }\textbf {\bibinfo {volume} {37}},\ \bibinfo {pages} {166}
  (\bibinfo {year} {2007})}\BibitemShut {NoStop}%
\bibitem [{\citenamefont {Hastings}(2021)}]{Has21}%
  \BibitemOpen
  \bibfield  {author} {\bibinfo {author} {\bibfnamefont {M.~B.}\ \bibnamefont
  {Hastings}},\ }\bibfield  {title} {\bibinfo {title} {The power of adiabatic
  quantum computation with no sign problem},\ }\href
  {https://doi.org/10.22331/q-2021-12-06-597} {\bibfield  {journal} {\bibinfo
  {journal} {Quantum}\ }\textbf {\bibinfo {volume} {5}},\ \bibinfo {pages}
  {597} (\bibinfo {year} {2021})}\BibitemShut {NoStop}%
\bibitem [{\citenamefont {Sanders}(2017)}]{San17}%
  \BibitemOpen
  \bibfield  {author} {\bibinfo {author} {\bibfnamefont {B.~C.}\ \bibnamefont
  {Sanders}},\ }\href {https://doi.org/10.1088/978-0-7503-1536-4} {\emph
  {\bibinfo {title} {How to Build a Quantum Computer}}},\ 2399-2891\ (\bibinfo
  {publisher} {IOP Publishing},\ \bibinfo {year} {2017})\BibitemShut {NoStop}%
\bibitem [{\citenamefont {Preskill}(2018)}]{Pre18}%
  \BibitemOpen
  \bibfield  {author} {\bibinfo {author} {\bibfnamefont {J.}~\bibnamefont
  {Preskill}},\ }\bibfield  {title} {\bibinfo {title} {Quantum computing in the
  {NISQ} era and beyond},\ }\href {https://doi.org/10.22331/q-2018-08-06-79}
  {\bibfield  {journal} {\bibinfo  {journal} {{Quantum}}\ }\textbf {\bibinfo
  {volume} {2}},\ \bibinfo {pages} {79} (\bibinfo {year} {2018})}\BibitemShut
  {NoStop}%
\bibitem [{\citenamefont {Bharti}\ \emph {et~al.}(2022)\citenamefont {Bharti},
  \citenamefont {Cervera-Lierta}, \citenamefont {Kyaw}, \citenamefont {Haug},
  \citenamefont {Alperin-Lea}, \citenamefont {Anand}, \citenamefont {Degroote},
  \citenamefont {Heimonen}, \citenamefont {Kottmann}, \citenamefont {Menke},
  \citenamefont {Mok}, \citenamefont {Sim}, \citenamefont {Kwek},\ and\
  \citenamefont {Aspuru-Guzik}}]{BCK+22}%
  \BibitemOpen
  \bibfield  {author} {\bibinfo {author} {\bibfnamefont {K.}~\bibnamefont
  {Bharti}}, \bibinfo {author} {\bibfnamefont {A.}~\bibnamefont
  {Cervera-Lierta}}, \bibinfo {author} {\bibfnamefont {T.~H.}\ \bibnamefont
  {Kyaw}}, \bibinfo {author} {\bibfnamefont {T.}~\bibnamefont {Haug}}, \bibinfo
  {author} {\bibfnamefont {S.}~\bibnamefont {Alperin-Lea}}, \bibinfo {author}
  {\bibfnamefont {A.}~\bibnamefont {Anand}}, \bibinfo {author} {\bibfnamefont
  {M.}~\bibnamefont {Degroote}}, \bibinfo {author} {\bibfnamefont
  {H.}~\bibnamefont {Heimonen}}, \bibinfo {author} {\bibfnamefont {J.~S.}\
  \bibnamefont {Kottmann}}, \bibinfo {author} {\bibfnamefont {T.}~\bibnamefont
  {Menke}}, \bibinfo {author} {\bibfnamefont {W.-K.}\ \bibnamefont {Mok}},
  \bibinfo {author} {\bibfnamefont {S.}~\bibnamefont {Sim}}, \bibinfo {author}
  {\bibfnamefont {L.-C.}\ \bibnamefont {Kwek}},\ and\ \bibinfo {author}
  {\bibfnamefont {A.}~\bibnamefont {Aspuru-Guzik}},\ }\bibfield  {title}
  {\bibinfo {title} {Noisy intermediate-scale quantum algorithms},\ }\href
  {https://doi.org/10.1103/RevModPhys.94.015004} {\bibfield  {journal}
  {\bibinfo  {journal} {Rev. Mod. Phys.}\ }\textbf {\bibinfo {volume} {94}},\
  \bibinfo {pages} {015004} (\bibinfo {year} {2022})}\BibitemShut {NoStop}%
\end{thebibliography}%
\end{document}